\documentclass[]{revtex4}
\usepackage{amsfonts}
\usepackage{amssymb,epsf}
\usepackage{graphicx}
\begin{document}

\title{Charged BTZ black holes in the context of massive gravity's rainbow}
\author{S. H. Hendi$^{1,2}$\footnote{
email address: hendi@shirazu.ac.ir}, S.
Panahiyan$^{1,3}$\footnote{ email address:
sh.panahiyan@gmail.com}, S. Upadhyay$^{4}$\footnote{ email
address: sudhakerupadhyay@gmail.com} and B. Eslam
Panah$^{1}$\footnote{ email address: behzad.eslampanah@gmail.com}
} \affiliation{$^1$ Physics Department and Biruni Observatory,
College of Sciences, Shiraz
University, Shiraz 71454, Iran\\
$^2$ Research Institute for Astronomy and Astrophysics of Maragha (RIAAM),
P.O. Box 55134-441, Maragha, Iran\\
$^3$ Physics Department, Shahid Beheshti University, Tehran 19839, Iran\\
$^4$ Centre for Theoretical Studies, Indian Institute of Technology
Kharagpur, Kharagpur-721302, WB, India}

\begin{abstract}
BTZ black holes are excellent laboratories for studying black hole
thermodynamics which is a bridge between classical general
relativity and quantum nature of gravitation. In addition,
three-dimensional gravity could have equipped us for exploring
some of the ideas behind the two dimensional conformal field
theory based on the $AdS_{3}/CFT_{2}$. Considering the significant
interests in these regards, we examine charged BTZ black holes. We
consider the system contains massive gravity with energy dependent
spacetime to enrich the results. In order to make high curvature
(energy) BTZ black holes more realistic, we modify the theory by
energy dependent constants. We investigate thermodynamic
properties of the solutions by calculating heat capacity and free
energy. We also analyze thermal stability and study the
possibility of Hawking-Page phase transition. At last, we study
geometrical thermodynamics of these black holes and compare the
results of various approaches.
\end{abstract}

\maketitle

\section{Introduction}

General relativity (GR) has been very successful in describing
different phenomena in low energy limits. Despite its success in
this regime, there are several issues which signal the necessity
of modifying this theory. Among them, one can name acceleration
expansion of the universe, existence of dark matter/dark energy
\cite{nj} and several other problems. GR predicts the existence of
massless property for gravitons as intermediate particles denoting
gravitational interactions. In order to solve the mentioned
problems, it has been proposed that Einstein gravity can be
modified to include massive gravitons. In other words, the
presence of massive and massless modes must be seen in
gravitational theory which is describing the system. This proposal
has been put into examination and its results were encouraging.
For example, the current acceleration of universe without
considering a
cosmological constant was explained by massive gravity \cite%
{kur,Expand1,Expand2}. In addition, it was shown that massive
spin-$2$ particles could be the candidate for dark matter since
their energy-momentum tensor behaves as that of dark matter fluid
\cite{Aoki,BG}. Furthermore, the solutions to hierarchy problem
point out the existence of massive modes, hence massive gravity
\cite{DvaliGP,DvaliGPI,DvaliG}. It is worthwhile to mention that
studies conducted in the context of string theory and quantum
gravity predict the presence of massive gravitons as well \cite%
{string1,string2,string3}.

The effects of massive gravity have been explored in the context
of astrophysical objects. For example, one can obtain a maximum
mass of neutron stars more than $3M_{sun}$ \cite{HBEslam} in the
context of massive gravity. In addition, it can modify the
thermodynamical quantities (behavior) of
black holes as well \cite%
{Bouchareb,Capela,Volkov,Babichev,Ghosh,ThermoMassive1,ThermoMassive2}.
Especially, the
existence of van der Waals like behavior for non-spherical black holes \cite%
{PRLwithMann}, remnant of temperature \cite{BTZMassive} and
anti-evaporation process for them were reported
\cite{anti1,anti2}. The possible effects of massive graviton on
gravitational waves produced during inflation were also studied
\cite{Gumrukcuoglu}.

Fierz and Pauli were the first researchers to start investigation
of a theory describing the possible free massive graviton
\cite{fr,fr1}. Later, it was found that this theory of massive
gravity suffers the Boulware-Deser (BD) ghost instability at the
non-linear level \cite{dg,dg1}. Recently, a significant progress
has been made toward constructing massive gravity theories without
such instability \cite{de}. Furthermore, the nonlinear massive
modifications to GR were also studied by many people in various
perspectives. Particularly, Refs. \cite{de,de1,kur} study a class
of nonlinear massive gravity theories in which the ghost field is
absent \cite{has,has1}. The simplest way to construct a massive
gravity is to simply add a mass term to the Einstein-Hilbert
action, giving the graviton a mass $m$ in such a way that GR is
recovered when $m\rightarrow 0$. Since a mass term breaks the
diffeomorphism invariance of the theories, hence, the energy
momentum is no longer conserved in this class of massive gravity.
In this paper, we employ a type of massive gravity which was
introduced by Vegh in Ref. \cite{Vegh}. This massive gravity has
specific applications in the context of lattice physics through
the concept of holography. Meaning that black hole solutions in
this gravity could have superconductor properties on the boundary
and massive gravitons could have lattice like behavior. While the
black hole solutions in the mentioned paper are obtained in
$4$-dimensions, here, we conduct our study in $3$-dimensions with
two other generalizations: energy dependent constants and
gravity's rainbow.

The usual energy-momentum relation or dispersion relation in
special relativity may be modified with corrections in the order
of Planck length by modifying the Lorentz--Poincar\'{e} symmetry.
This deformed formalism of special relativity is known as "Doubly
Special Relativity" \cite{ame,ame1,mag,kow}. The generalization of
this idea to curved spacetime was done by Magueijo and Smolin
\cite{mag1}. This formalism is known commonly as gravity's
rainbow. The idea of gravity's rainbow formalism is that the free
falling observers who make measurements with energy $E$\ will
observe the same laws of physics as in modified special
relativity. In fact, the gravity's rainbow produces a correction
to the spacetime metric which becomes significant as soon as the
particle's energy/momentum approaches the Planck energy. In this
formalism, the connection and curvature depend on energy in such a
way that the usual Einstein's equations is replaced by a one
parameter family of equations. In this context, the Gauss-Bonnet
and dilaton gravities were generalized to energy-dependent
Gauss-Bonnet and dilatonic
theories of gravity and their black hole solutions were studied \cite%
{fai,faifai}. Recently, the critical behavior of uncharged and
charged black holes in Gauss-Bonnet gravity's rainbow was analyzed
and it was found that the generalization to a charged case puts an
energy dependent restriction on different parameters \cite{fai1}.
Two classes of $F(R)$ gravity's rainbow solutions were also
investigated \cite{hend00}. In the first case, the energy
dependent $F(R)$\ gravity without energy momentum tensor was
studied and, secondly, $F(R)$\ gravity's rainbow in the presence
of conformally invariant Maxwell source was analyzed. In addition,
the Starobinsky model of inflation in the context of gravity's
rainbow was investigated where rainbow functions are written in
the power-law form of the Hubble parameter \cite{aut}. In this
context, the spectral index of curvature perturbation, the
tensor-to-scalar ratio and consistency of these models with Planck
2015 data are also discussed. Moreover, Galileon
gravity's rainbow by considering Vaidya spacetime has been studied in \cite%
{Vaidya}. Also, the Unruh, Hawking, fiducial and free-fall temperatures of
the black hole in gravity's rainbow have been investigated in Refs. \cite%
{Bibhas1,Gim}. The absence of black holes at LHC \cite{AliPLB},
remnants of black objects \cite{AliJHEP}, nonsingular universes in
Einstein and Gauss-Bonnet gravities \cite{AliHendi1,AliHendi2}
have been analyzed in the gravity's rainbow background. From
astrophysical perspective, it was shown that the existence of
energy dependent spacetime can modify the hydrostatic equilibrium
equation of stars \cite{HendiJCAP}. In addition, the modifications
on Hawking-Page phase transition \cite{HW1,HW2}, wave function of
the universe \cite{Wave} and generalization of black hole
thermodynamics \cite{Mod} are investigated in the context of
gravity's rainbow.

To explore the foundations of classical and quantum gravity, GR in $3$%
-dimensions has become a very popular model \cite{car}. Of the
drawbacks of the GR model in three dimensions were that there were
no Newtonian limit \cite{jd} and no propagating degrees of
freedom. In 1992, Ba\~{n}ados, Teitelboim and Zanelli (BTZ) came
with surprising result that three dimensional gravity with a
negative cosmological constant has a black hole solution
\cite{btz}. BTZ black holes provide a good understanding of
certain central issues like black hole thermodynamics
\cite{car1,ast,sar}, quantum gravity, string and gauge theory and
more importantly the AdS/CFT conjecture \cite{wit,car2}.
Furthermore, BTZ solutions perform a crucial role in improving our
perception of gravitational interaction in low dimensional
spacetime \cite{wit1}. The charged BTZ black hole is the analogous
solution of adS-Maxwell gravity in three dimensions
\cite{car1,mar,cle}. Recently, thermodynamics and phase structure
of the charged black hole solutions in both grand canonical and
canonical ensembles were studied
\cite{CaiMassive,hend1,BTZMassive}. Furthermore, thermodynamical
phase transition of BTZ black holes through the Landau-Lifshitz
theory \cite{Purt} and quantum correction of the entropy in the
noncommutative BTZ black holes \cite{correction} have been
investigated.

As we mentioned before, the core stone of the gravity's rainbow is
doubly special relativity (DSR). In fact, the gravity's rainbow in
its first proposal was introduced as "doubly general relativity"
\cite{mag1}. Therefore, in order to outline the properties of the
gravity's rainbow in three dimensions, one should regard the DSR
in three dimensions. Specifically speaking, in a pioneering work
by Freidel, \emph{et al} \cite{Freidel}, it was shown that gravity
in $2+1$ dimensions coupled to point particles results into a
nontrivial example of DSR. Therefore, it is stated that (quantum)
gravity in $2+1$ dimensions coupled to point particles is indeed
just a DSR theory. This point was shown by the fact that symmetry
algebra of quantum gravity in $2+1$ dimensions is not
Poincar\'{e}, but it is a (quantum) $\kappa $-deformed
Poincar\'{e} \cite{Welling}. On the other hand, the symmetry
algebra of a DSR theory is also $\kappa $-deformed Poincar\'{e}.
It is possible to explicitly map the phase space of quantum
gravity in $2+1$ coupled to a single point particle to the algebra
symmetry generators of a DSR theory. In addition, in another
pioneering work, Blaut, \textit{et al} showed that depending on
the direction of deformation of $\kappa $-deformed Poincar\'{e},
phase spaces of single particle in DSR theories have the
energy-momentum spaces of the form of de Sitter, anti de Sitter
and flat space \cite{Blaut}. The study was conducted for arbitrary
dimensions including $3$-disunions. Now, remembering that
gravity's rainbow essentially is a generalization of the DSR,
therefore, it is expected that it preserves the fundamental
properties of the DSR which enables us to recognize the origin of
the gravity's rainbow and the presence of its effects.

In addition, following the same method of Ref. \cite{Assanioussi},
one can find that the effective metric of a quantum cosmological
model describing the emergent spacetime in arbitrary dimensions
should be indeed of the rainbow type, without needing to any
ad-hoc input. Nevertheless, in this paper, we take into account
the three dimensional energy dependent spacetime as a toy model
and investigate its nontrivial thermodynamic properties caused by
energy functions.

Furthermore, it is worthwhile to mention that in $2+1$ dimensions,
the Planck energy is defined as $E_{P}^{\left( 2+1\right)
}=c^{4}/G^{\left( 2+1\right) }$ in which $c$ is the speed of light
and $G$ is gravitational constant in $2+1$ dimensions. Thus, the
dimension of $G$ is inverse mass and therefore, it may seem that
theory under consideration is a classical one, but that is not the
case. The existence of matter in $2+1$ gravity causes the geometry
of the spacetime to be conical one with specific deformed
asymptotic conditions which depend on $G$.  This deformation has
specific effects on algebra of the classical phase which
highlights yet another reason why $2+1$ gravity is a DSR theory.
The fact is $G$ here is identified with inverse of the $\kappa$
deformation parameter of the (quantum) $\kappa $-Poincar\'{e}
algebra. As it was pointed out, essentially, the $3$-dimensional
quantum gravity coupled with a point particle is a DSR theory. In
Ref. \cite{Kowalski-Glikman}, the relation between $2+1$ quantum
gravity and DSR was described in details. Livine, \textit{et al}
in their work studied quantum geometry of a $3$-dimensional DSR
\cite{Livine}. For furthers studies regarding the quantum
applications of the DSR theory in $3$-dimensions, we refer the
readers to Refs. \cite{q1,q2,q3,q4}. This shows that DSR theory
could be a quantum one. Since the gravity's rainbow is a
generalization of the DSR, we expect that it preserves its
properties as well which indicates that the effects of gravity's
rainbow could be quantum-like ones. But we should emphasize it
that here, our focus and main motivations are concerning the
effects of gravity's rainbow alongside of massive gravity on
semi-classical thermodynamical behavior of the black holes in
three dimensions.

Geometrical thermodynamics (GTs) is one of the interesting methods
for studying the properties of thermodynamical systems. In this
method, the Riemannian geometry is used to construct phase space.
The Ricci scalar of this phase space is employed to extract some
information regarding thermodynamical behavior of the system. In
other words, GTs is a bridge between geometry and thermodynamics.
The geometrical information of Ricci scalar are obtained through
its divergencies. These divergencies determine three important
points;

I) Bound points which separate solutions with positive temperature (physical
systems) from those with negative temperature (non-physical systems).

II) Phase transition points which represent discontinuities in
thermodynamical quantities such as heat capacity.

III) The sign of Ricci scalar around divergence points determines
the nature of interaction on molecular level \cite{Rupp}.

Weinhold introduced the first geometrical thermodynamical approach in 1975.
Weinhold's approach was based on internal energy as thermodynamical
potential \cite{WeinholdI,WeinholdII}. Then, Ruppeiner proposed an
alternative approach which has entropy as its thermodynamical potential \cite%
{RuppeinerI,RuppeinerII}. Since Weinhold and Ruppeiner's approaches are not
Legendre invariant, Quevedo introduced another approach for GTs \cite%
{QuevedoI,QuevedoII}. Several investigations regarding thermodynamics of the
black holes through these methods were done in Refs. \cite%
{HanC,BravettiMMA,Ma,GarciaMC,ZhangCY,MoLW2016,Sanchez,Soroushfar1}.

On the other hand, it was shown that mentioned methods may
confront specific problems in describing thermodynamical
properties of the black holes (see refs.
\cite{HPEMI,HPEMII,HPEMIII,HPEMIV}, for more details). In other
words, obtained results of these three approaches were not
consistent with those extracted from other methods. Therefore, in
order to remove the shortcomings of other methods, Hendi et al
proposed a new thermodynamical metric (HPEM) \cite{HPEMI}. It is
notable that, it was shown that employing this new metric leads
into consistent results regarding thermodynamical properties of
the black holes. We refer the reader to Ref. \cite{Wen} for a
comparative study regarding these four thermodynamical metrics and Ref. \cite%
{zhang} regarding the application of HPEM metric in studying critical
behavior of the system.

The paper at hand, regards three dimensional charge black holes
with three generalizations; energy dependent constants, gravity's
rainbow and massive gravity. Recently, three dimensional charged
black holes in the presence of massive gravity have been
investigated \cite{BTZMassive}. Here, we apply the generalization
to gravity's rainbow to understand how this generalization would
modify previous results. In fact, we would like to see how the
energy dependent spacetime would affects thermodynamical structure
of the massive charged BTZ black holes. Such generalization is
necessary from different aspects; First of all, black holes and
their physics are governed by high energy physics. This indicates
that it is necessary to include the upper limit of Plank energy on
energies that particles can acquire. This is the prescription of
gravity's rainbow. On the other hand, it is stated that quantum
corrections of quantum gravity could be observed as energy
dependency of the spacetime \cite{quantum1,quantum2}. In other
words, one could include the quantum corrections in form of the
energy dependency of spacetime which leads to gravity's rainbow.
These provide us with motivations to consider gravity's rainbow
alongside of massive gravity.

The consideration of energy dependency of the constants is rooted
in studies that are conducted in the context of renormalization
group flow \cite{flow}. These studies emphasized on the energy
dependency of constants on the scale of theory probed. Through
several studies, the flow of cosmological \cite{cosmflow} and
Newton \cite{newtonflow} constants were examined. Since the scale
measurement of theory under consideration depends on the energy
that probe can acquire, therefore, it is logical to consider all
the constants as energy dependent ones. Such consideration has
been taken into account in the context of Gauss-Bonnet gravity and
it was shown that it enriches both geometrical and thermodynamical
aspects of the black holes \cite{faifai,fai1}. Here too, we employ
such consideration to take all the constants energy dependent.
Such an idea provides different perspectives for the observers who
are at different distance from black holes under consideration. In
addition, it would have specific contributions to other studies
that could be conducted in the context of these types of the black
holes (we refer the reader to Refs. \cite{suresh,prasia} for some
examples). Our other motivation for considering such set up for
black holes is to provide a number of generalizations. These
specific generalizations are effective in specific regions of
energy which provide a better picture regarding the nature of
black holes.

The outline of paper is as follows. First, we will introduce the
basic field equations and metric, and extract black holes
solutions. Next, thermodynamical quantities are calculated and the
first law of thermodynamics for black holes is examined. Then
thermodynamical properties of the black holes are studied through,
mass, temperature, heat capacity and free energy. Next, we will
study thermodynamics of these black holes in the context of GTs
and show the consistency of its results with divergencies and
bound points of the heat capacity. The paper is finished with some
closing remarks.

\section{Black hole solutions}

\label{FieldEq}

The general formalism of gravity's rainbow could be obtained by using a
deformation of the standard energy-momentum relation
\begin{equation}
E^{2}f^{2}(\varepsilon )-p^{2}g^{2}(\varepsilon )=m^{2},  \label{MDR}
\end{equation}%
where the dimensionless energy ratio is $\varepsilon =E/E_{P}$ in which $E$
and $E_{P}$ are, respectively, the energy of test particle and the Planck
energy. Since the energy of a test particle can not exceed the Plank energy,
we should remind $0<\varepsilon \leq 1.$ Here, $f(\varepsilon )$ and $%
g(\varepsilon )$ are energy functions which are restricted with the
following condition in infrared limit
\begin{equation}
\lim\limits_{\varepsilon \rightarrow 0}f(\varepsilon )=1,\qquad
\lim\limits_{\varepsilon \rightarrow 0}g(\varepsilon )=1.
\end{equation}%
Regarding the analogy between the energy-momentum four vector
$(E,\vec{p})$ with time-space one $(t,\vec{x})$, it is possible to
use the energy functions to build an energy dependent spacetime
with following recipe
\begin{equation}
\hat{g}(\varepsilon )=\eta ^{ab}e_{a}(\varepsilon )\otimes e_{b}(\varepsilon
),  \label{rainmetric}
\end{equation}%
where
\begin{equation}
e_{0}(\varepsilon )=\frac{1}{f(\varepsilon )}\tilde{e}_{0},\qquad
e_{i}(\varepsilon )=\frac{1}{g(\varepsilon )}\tilde{e}_{i},
\end{equation}%
with $\tilde{e}_{0}$ and $\tilde{e}_{i}$ being the energy independent frame
fields (the algorithm of (\ref{rainmetric}) may be originate from the
analogy between two invariant relations; the energy-momentum relation and
the line element invariant).

The $3$-dimensional form of massive gravity's Lagrangian is
\[
L_{massive}=m\left( \varepsilon \right) ^{2}\sum_{i=1}^{3}c_{i}(\varepsilon )%
\mathcal{U}_{i}(g,f),
\]%
in which $c(\varepsilon )_{i}$'s are some energy dependent constants and $%
\mathcal{U}_{i}$'s are symmetric polynomials of the eigenvalues of the $%
3\times 3$ matrix $\mathcal{K}_{\nu }^{\mu }=\sqrt{g^{\mu \alpha
}f_{\alpha \nu }}$, which can be written as follows
\begin{eqnarray}
\mathcal{U}_{1} &=&\left[ \mathcal{K}\right] ,\;\;\;\;\;\mathcal{U}_{2}=%
\left[ \mathcal{K}\right] ^{2}-\left[ \mathcal{K}^{2}\right] ,\;\;\;\;\;%
\mathcal{U}_{3}=\left[ \mathcal{K}\right] ^{3}-3\left[ \mathcal{K}\right] %
\left[ \mathcal{K}^{2}\right] +2\left[ \mathcal{K}^{3}\right] ,
\end{eqnarray}
where by using variational principle, we could have
\begin{eqnarray}
\chi _{\mu \nu } &=&-\frac{c_{1}(\varepsilon )}{2}\left( \mathcal{U}%
_{1}g_{\mu \nu }-\mathcal{K}_{\mu \nu }\right) -\frac{c_{2}(\varepsilon )}{2}%
\left( \mathcal{U}_{2}g_{\mu \nu }-2\mathcal{U}_{1}\mathcal{K}_{\mu \nu }+2%
\mathcal{K}_{\mu \nu }^{2}\right)   \nonumber \\
&&-\frac{c_{3}(\varepsilon )}{2}(\mathcal{U}%
_{3}g_{\mu \nu }-3\mathcal{U}_{2}\mathcal{K}_{\mu \nu
}+6\mathcal{U}_{1}\mathcal{K}_{\mu \nu }^{2}-6\mathcal{K}_{\mu \nu
}^{3}). \label{massiveTerm}
\end{eqnarray}

The only non-zero term of massive gravity is $\mathcal{U}_{1}$
while the other higher order terms are vanished. By taking this
fact into account, one can find following action governing our
black holes of interest
\begin{equation}
\mathcal{I}=-\frac{1}{16\pi G(\varepsilon )}\int d^{3}x\sqrt{-g}\left[
\mathcal{R}-2\Lambda \left( \varepsilon \right) -\mathcal{F}+m\left(
\varepsilon \right) ^{2}c_{1}(\varepsilon )\mathcal{U}_{1}(g,f)\right] ,
\label{Action}
\end{equation}%
where $\mathcal{R}$ and $\mathcal{F}$ are, respectively, the scalar
curvature and the Lagrangian of Maxwell electrodynamics, $G(\varepsilon )$
is the energy dependent gravitational constant, $\Lambda \left( \varepsilon
\right) $ is the energy dependent cosmological constant and $f$ is an energy
dependent fixed symmetric tensor. In addition, $\mathcal{F}=F_{\mu \nu
}F^{\mu \nu }$\ is the Maxwell invariant, in which $F_{\mu \nu }=\partial
_{\mu }A_{\nu }-\partial _{\nu }A_{\mu }$ is the Faraday tensor with $A_{\mu
}$ as the gauge potential. Taking the action (\ref{Action}) into account and
using the variational principle, we obtain the field equations corresponding
to the gravitation and gauge fields as
\begin{equation}
R_{\mu \nu }-\left( \frac{R}{2}-\Lambda \left( \varepsilon \right) \right)
g_{\mu \nu }+G\left( \varepsilon \right) \left( \frac{1}{2}g_{\mu \nu }%
\mathcal{F}-2L_{\mathcal{F}}F_{\mu \rho }F_{\nu }^{\rho }\right) +m\left(
\varepsilon \right) ^{2}\chi _{\mu \nu }=0,  \label{Field
equation}
\end{equation}%
\begin{equation}
\partial _{\mu }\left( \sqrt{-g}F^{\mu \nu }\right) =0.
\label{Maxwell equation}
\end{equation}

Here, we are interested in static charged black hole solutions,
and therefore, we consider the metric of $3$-dimensional spacetime
with the following energy dependent line element
\begin{equation}
ds^{2}=-\frac{\psi (r)}{f(\varepsilon )^{2}}dt^{2}+\frac{1}{%
g(\varepsilon )^{2}}\left( \frac{dr^{2}}{\psi (r)}+r^{2}d\varphi
^{2}\right) ,  \label{metric}
\end{equation}%
in which $\psi (r)$ is the metric function of our black holes.

Our main motivation is to obtain massive black holes in the
context of gravity's rainbow. This requires specific modifications
in the reference metric in form of
\begin{equation}
f_{\mu \nu }=diag\left( 0,0,\frac{c(\varepsilon )^{2}}{g(\varepsilon )^{2}}%
\right) ,  \label{f11}
\end{equation}%
where $c(\varepsilon )$ is an arbitrary energy dependent positive
constant. This choice of reference metric is motivated from
holographical perspective of strongly interacting quantum field
theories. Vegh, in his work showed that this choice of reference
metric provides the possibility of the graviton to have lattice
like behaviour by showing a Drude peak which approaches a delta
function in the massless gravity limit \cite{Vegh}. Using this
metric ansatz (\ref{f11}), $\mathcal{U}_{1}$ will be calculated in
form of \cite{CaiMassive}
\begin{equation}
\mathcal{U}_{1}=\frac{c(\varepsilon )}{r}.  \label{U}
\end{equation}

In order to have a radial electric field, we consider the following gauge
potential%
\begin{equation}
A_{\mu }=h(r) \delta _{\mu }^{t},  \label{gauge potential}
\end{equation}%
where by using the metric (\ref{metric}) with the Maxwell field equation (%
\ref{Maxwell equation}), one can find the following differential
equation
\begin{equation}
h^{\prime }(r)+rh^{\prime \prime }(r)=0,  \label{heq}
\end{equation}%
in which the prime and double prime are representing the first and second
derivatives with respect to $r$, respectively. It is a matter of calculation
to solve Eq. (\ref{heq}), yielding
\begin{equation}
h(r)=q\left( \varepsilon \right) \ln \left(\frac{r}{l(\varepsilon
)}\right), \label{h(r)}
\end{equation}%
where $q(\varepsilon )$ is an energy dependent integration constant related
to the electric charge and $l(\varepsilon )$ is an arbitrary energy
dependent constant with length dimension which is considered for the sake of
having dimensionless logarithmic argument. It is worthwhile to mention that
the corresponding electromagnetic field tensor is $F_{tr}=\frac{%
q(\varepsilon )}{r}$, which is independent of $l(\varepsilon )$.

In order to obtain metric function, $\psi (r)$, we use Eq.
(\ref{Field equation}) with Eq. (\ref{metric}), and obtain the
following differential equations
\begin{eqnarray}
&&rg(\varepsilon )^{2}\psi ^{\prime }(r)+2r^{2}\Lambda \left(
\varepsilon \right) +2G(\varepsilon )g(\varepsilon
)^{2}f(\varepsilon )^{2}q\left( \varepsilon \right) ^{2}-m\left(
\varepsilon \right) ^{2}c(\varepsilon )c_{1}(\varepsilon )r=0,
\label{eqENMax1} \\
&&\frac{r^{2}}{2}g(\varepsilon )^{2}\psi ^{\prime \prime
}(r)+\Lambda \left( \varepsilon \right) r^{2}-G(\varepsilon
)g(\varepsilon )^{2}f(\varepsilon )^{2}q\left( \varepsilon \right)
^{2}=0,  \label{eqENMax2}
\end{eqnarray}%
which correspond to $tt$ (or $rr$) and $\varphi \varphi $ components of Eq. (%
\ref{Field equation}), respectively. It is straightforward to show that
metric function is obtained as%
\begin{equation}
\psi (r)=-\frac{\Lambda \left( \varepsilon \right) r^{2}}{%
g(\varepsilon )^{2}}-m_{0}\left( \varepsilon \right) -2G\left(
\varepsilon \right) f(\varepsilon )^{2}q\left( \varepsilon \right)
^{2}\ln \left(\frac{r}{l(\varepsilon )}\right)+\frac{m\left(
\varepsilon \right) ^{2}c(\varepsilon )c_{1}(\varepsilon
)r}{g(\varepsilon )^{2}},  \label{f(r)ENMax}
\end{equation}%
where $m_{0}(\varepsilon )$ is an energy dependent integration
constant related to the total mass of black holes. It is
worthwhile to mention that the resulting metric function
(\ref{f(r)ENMax}) satisfies all the components of field equation
(\ref{Field equation}), simultaneously. In the absence of
massive parameter (i.e. $m(\varepsilon )=0$), the metric function Eq. (\ref%
{f(r)ENMax}) will be reduced to
\begin{equation}
\psi (r)=-\frac{\Lambda \left( \varepsilon \right) r^{2}}{%
g(\varepsilon )^{2}}-m_{0}\left( \varepsilon \right) -2G\left(
\varepsilon \right) f(\varepsilon )^{2}q\left( \varepsilon \right)
^{2}\ln \left(\frac{r}{l(\varepsilon )}\right).
\end{equation}

Our next step is examination of the geometrical structure of solutions.
First, we should look for the existence of essential singularity(ies). The
Ricci and Kretschmann scalars of the solutions are, respectively,
\begin{eqnarray}
R &=&6\Lambda \left( \varepsilon \right) +\frac{2G\left( \varepsilon\right)
f(\varepsilon )^{2}q\left( \varepsilon \right) ^{2}}{r^{2}}-\frac{2m\left(
\varepsilon \right) ^{2}c(\varepsilon )c_{1}(\varepsilon )}{r}, \\
R_{\alpha \beta \gamma \delta }R^{\alpha \beta \gamma \delta } &=&12\Lambda
\left( \varepsilon \right) ^{2}-\frac{8\Lambda \left( \varepsilon \right)
m\left( \varepsilon \right) ^{2}c(\varepsilon )c_{1}(\varepsilon )}{r}+\frac{%
2\left[ m\left( \varepsilon \right) ^{4}c(\varepsilon )^{2}c_{1}(\varepsilon
)^{2}+4G\left( \varepsilon \right) g\left( \varepsilon \right)
^{2}f(\varepsilon )^{2}\Lambda \left( \varepsilon \right) q\left(
\varepsilon \right) ^{2}\right] }{r^{2}}  \nonumber \\
&-&\frac{8G\left( \varepsilon \right) g\left( \varepsilon \right)
^{2}f(\varepsilon )^{2}q\left( \varepsilon \right) ^{2}m\left( \varepsilon
\right) ^{2}c(\varepsilon )c_{1}(\varepsilon )}{r^{3}}+\frac{12G\left(
\varepsilon \right) ^{2}g\left( \varepsilon \right) ^{4}f(\varepsilon
)^{4}q\left( \varepsilon \right) ^{4}}{r^{4}}.
\end{eqnarray}

These relations confirm that there is an essential curvature singularity at $%
r=0$. For the limit of $r\longrightarrow \infty $, the Ricci and Kretschmann
scalars yield the values $6\Lambda \left( \varepsilon \right) $ and $%
12\Lambda \left( \varepsilon \right) ^{2}$ , respectively, which show that
for $\Lambda \left( \varepsilon \right) >0$ ($\Lambda \left( \varepsilon
\right) <0$), the asymptotical behavior of the solution is (a)dS with an
energy dependent cosmological constant.

\begin{figure}[tbp]
$%
\begin{array}{ccc}
\epsfxsize=5cm \epsffile{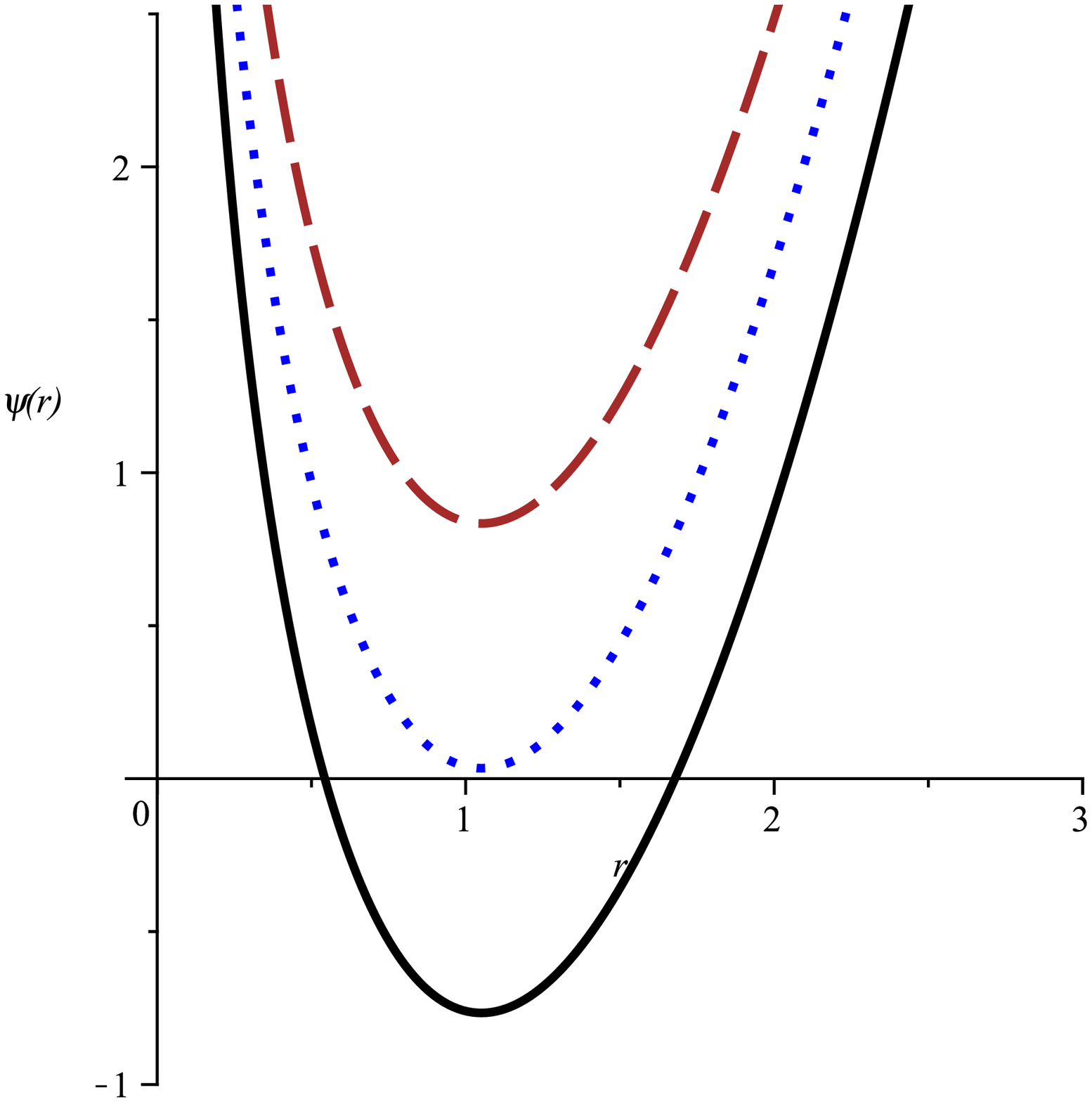} & \epsfxsize=5cm %
\epsffile{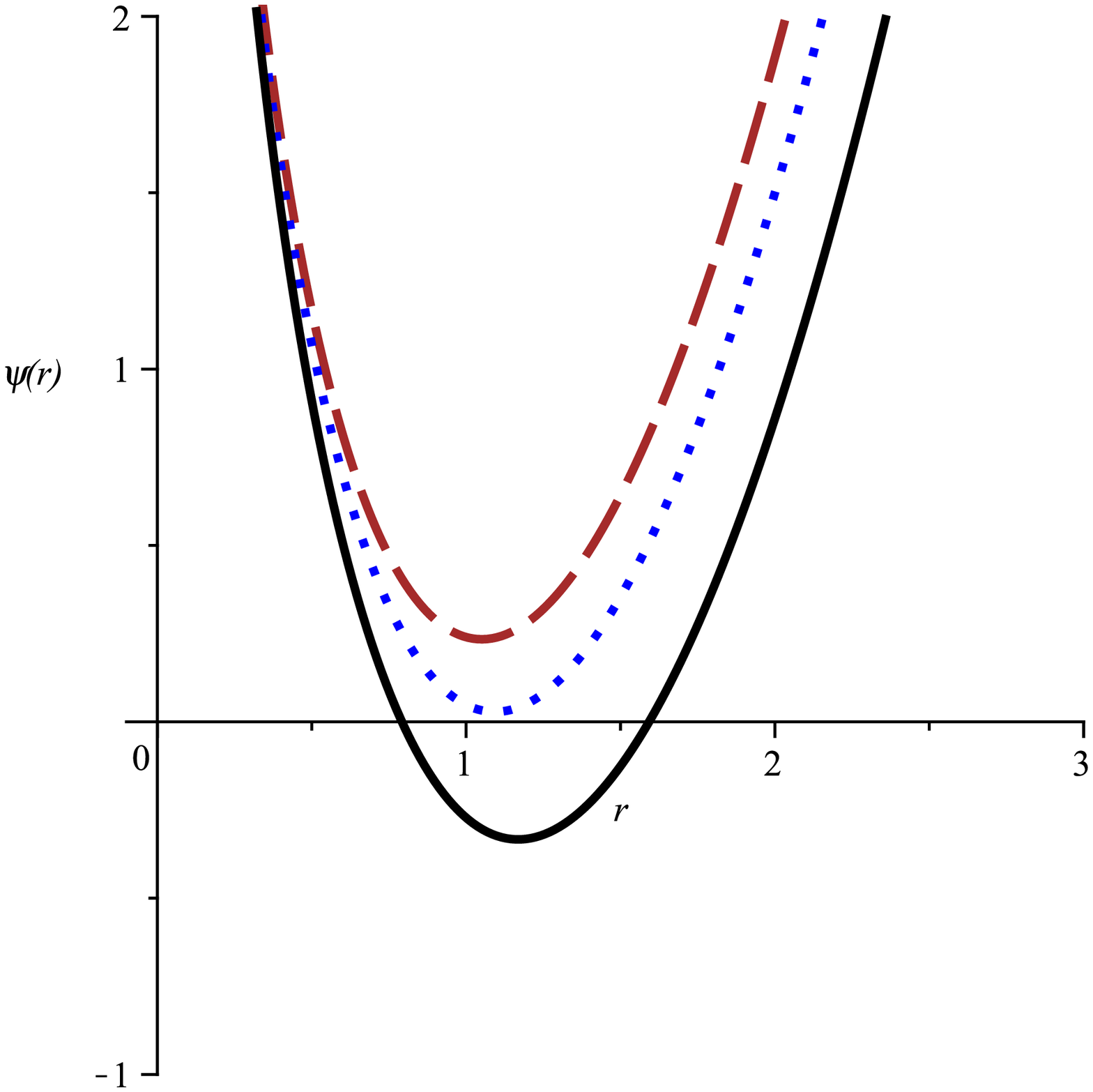} & \epsfxsize=5cm \epsffile{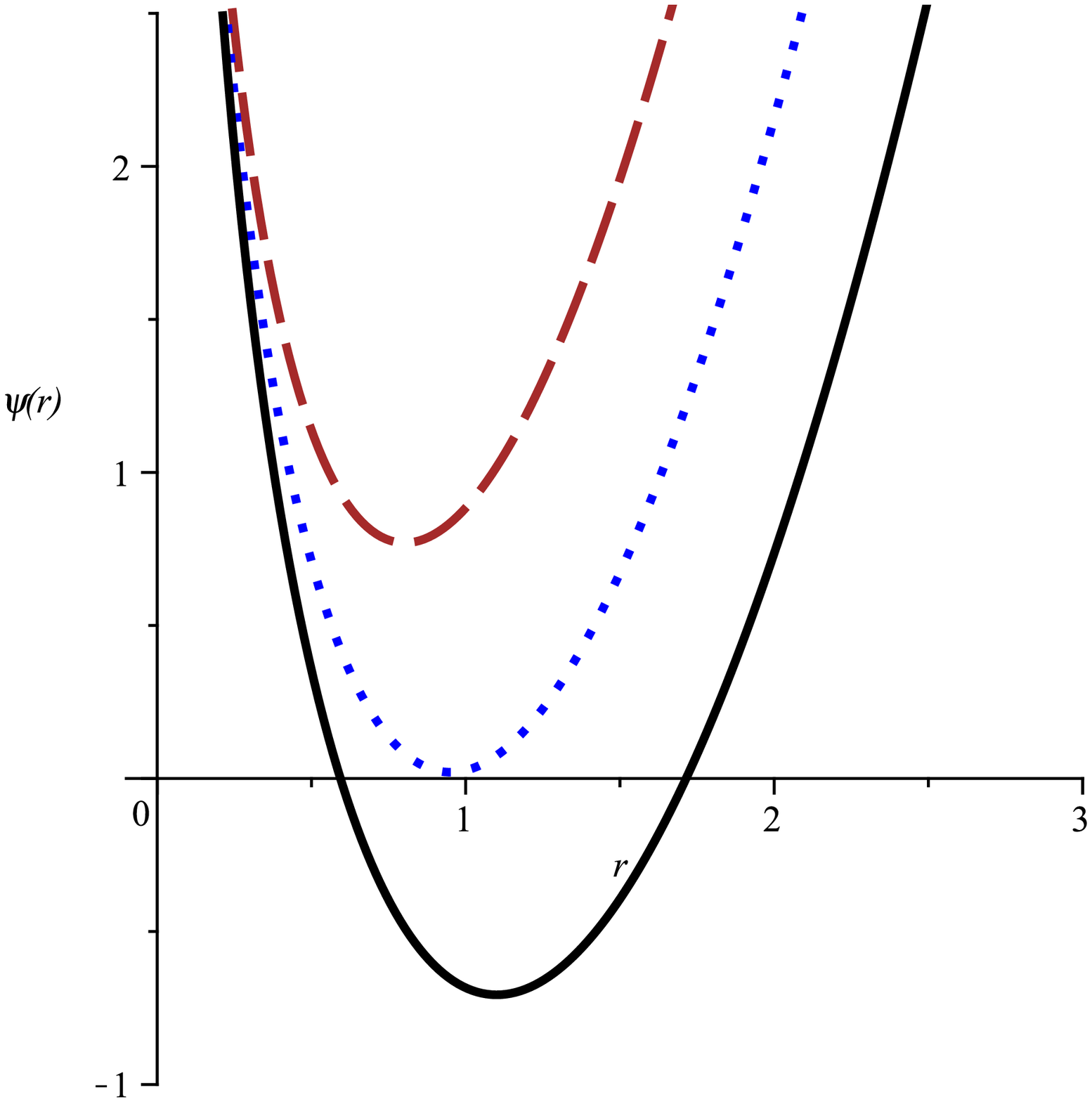}%
\end{array}
$%
\caption{$\protect\psi (r)$ versus $r$ for $\Lambda (%
\protect\varepsilon )=-1$, $G(\protect\varepsilon )=l(\protect\varepsilon %
)=1 $, $q(\protect\varepsilon )=1.5$, $g(\protect\varepsilon )=1$, $f(%
\protect\varepsilon )=0.8 $ and $m(\protect\varepsilon )=0.8$. \newline
Left panel: $c(\protect\varepsilon )=c_{1}(\protect\varepsilon )=1$, $m_{0}(%
\protect\varepsilon )=0.8$ (dashed line), $m_{0}(\protect\varepsilon )=1.6$
(dotted line) and $m(\protect\varepsilon )=2.4$ (continuous line). \newline
Middle panel: $m_{0}(\protect\varepsilon )=1.4$, $c(\protect\varepsilon )=1$%
, $c_{1}(\protect\varepsilon )=1$ (dashed line), $c_{1}(\protect\varepsilon %
)=0.7$ (dotted line) and $c_{1}(\protect\varepsilon )=0.2$ (continuous
line). \newline
Right panel: $m_{0}(\protect\varepsilon )=2.1$, $c_{1}(\protect\varepsilon %
)=1$, $c(\protect\varepsilon )=3.1$ (dashed line), $c(\protect\varepsilon %
)=1.76$ (dotted line) and $c(\protect\varepsilon )=0.65$ (continuous line).}
\label{Fig1}
\end{figure}

Our final step in this section is investigation of other
geometrical properties such as the existence of regular horizon.
For this purpose, we have plotted Fig. \ref{Fig1} to find the real
positive roots of metric function. Evidently, depending on the
choices of different parameters, it is possible to observe, two
horizons, one extreme horizon and without horizon (naked
singularity) for these solutions (see Fig. \ref{Fig1} for more
details). This confirms that the singularity can be covered with
an event horizon, and therefore, our solutions are basically
representing black holes.

\subsection{Thermodynamics}

Now, we intend to calculate the conserved and thermodynamic quantities of
the solutions and examine the validity of the first law of thermodynamics.

Using the standard definition of Hawking temperature with its
relation to the surface gravity on the outer horizon $r_{+}$, we
obtain
\begin{equation}
T=-\frac{\Lambda \left( \varepsilon \right) r_{+}}{2\pi f\left( \varepsilon
\right) g\left( \varepsilon \right) }+\frac{m\left( \varepsilon \right)
^{2}c(\varepsilon )c_{1}(\varepsilon )}{4\pi f\left( \varepsilon \right)
g\left( \varepsilon \right) }-\frac{f\left( \varepsilon \right) g\left(
\varepsilon \right) G\left( \varepsilon \right) q\left( \varepsilon \right)
^{2}}{2\pi r_{+}}.  \label{TotalTT}
\end{equation}

Furthermore, calculating the flux of electric field at infinity and using
the Gauss's law, we can compute the electric charge, $Q$, as
\begin{equation}
Q=\frac{1}{2}f\left( \varepsilon \right) G\left( \varepsilon \right) q\left(
\varepsilon \right) .  \label{TotalQ}
\end{equation}

Since we are working in Einstein gravity, the entropy of black holes can be
obtained by employing the area law. According to this law, we can derive the
entropy as a quarter of event horizon area \cite%
{hawking1,hawking2,hawking3,Hunter1,Hunter2,Hunter3}
\begin{equation}
S=\frac{\pi }{2g\left( \varepsilon \right) }r_{+}.  \label{TotalS}
\end{equation}

Also, we can obtain the total mass of solutions by using the
Hamiltonian approach and/or the counterterm method with the
following explicit form
\begin{equation}
M=\frac{m_{0}\left( \varepsilon \right) }{8f\left( \varepsilon \right) },
\label{TotalM}
\end{equation}
where $m_{0}\left( \varepsilon \right) $ can be computed from the metric
function (\ref{f(r)ENMax}) on the horizon ($\psi \left(
r=r_{+},\varepsilon\right) =0$), and consequently, it may be presented with
the following expression
\begin{equation}
m_{0}\left( \varepsilon \right) =-\frac{\Lambda (\varepsilon )r_{+}^{2}}{%
g(\varepsilon )^{2}}-2G(\varepsilon)f(\varepsilon
)^{2}q(\varepsilon )^{2} \ln \left( \frac{r_{+}}{l(\varepsilon
)}\right) +\frac{ m(\varepsilon )^{2}c(\varepsilon
)c_{1}(\varepsilon )r_{+}}{g(\varepsilon )^{2}}.  \nonumber
\end{equation}

The electric potential, $U$, is calculated through the difference of gauge
potential between the reference and the horizon, and it is given as
\begin{equation}
U=A_{\mu }\chi ^{\mu }\left\vert _{r\rightarrow reference}\right. -A_{\mu
}\chi ^{\mu }\left\vert _{r\rightarrow r_{+}}\right. =-q\left( \varepsilon
\right) \ln \left( \frac{r_{+}}{l(\varepsilon )}\right) .  \label{TotalU}
\end{equation}

Now, we are in a position to check the validity of the first law of
thermodynamics. Exploiting thermodynamic quantities such as electric charge (%
\ref{TotalQ}), entropy (\ref{TotalS}) and mass (\ref{TotalM}), with the
first law of black hole thermodynamics
\begin{equation}
dM=TdS+UdQ,
\end{equation}%
one can define the intensive parameters conjugate to $S$ and $Q$. These
quantities are the temperature and the electric potential
\begin{equation}
T=\left( \frac{\partial M}{\partial S}\right) _{Q}\ \ \ \mbox{and} \ \ \ \ \
\ \ \ \ U=\left( \frac{\partial M}{\partial Q}\right) _{S}.  \label{TU}
\end{equation}

Using the obtained electric charge (\ref{TotalQ}) and entropy (\ref{TotalS})
with total mass of the black holes (\ref{TotalM}), one can find following
Smarr-type formula
\begin{equation}
M\left( S,Q\right) =-\,{\frac{\Lambda \left( \varepsilon \right) {S}^{2}}{%
2f\left( \varepsilon \right) {\pi }^{2}}}+\,{\frac{m\left( \varepsilon
\right) ^{2}S\,c(\varepsilon )c_{1}(\varepsilon )}{4g\left( \varepsilon
\right) f\left( \varepsilon \right) \pi }}-\frac{{Q}^{2}}{G\left(
\varepsilon \right) f\left( \varepsilon \right) }\ln \left( 2\,{\frac{%
Sg\left( \varepsilon \right) }{\pi \,l(\varepsilon )}}\right) .
\label{MSmarr}
\end{equation}

It is a matter of calculation to show that the calculated temperature and
electric potential by using Eq. (\ref{TU}) are same as those calculated for
the temperature (\ref{TotalTT}) and the electric potential (\ref{TotalU}).
In other word, although massive term and gravity's rainbow modify some of
thermodynamic quantities, the first law of thermodynamics is still valid.

Our next thermodynamical quantity of the interest is the heat capacity. This
quantity contains information regarding the phase transition points and
conditions for thermal stability. The stability of solutions is governed by
the sign of heat capacity; its positivity indicates thermal stability while
the opposite represents instability. The phase transition points are
extracted by finding the divergencies of the heat capacity. In other words,
divergencies of the heat capacity may be characterized with a second order
phase transitions. In addition, since the roots of heat capacity and
temperature are the same (due to form of the heat capacity), the roots of
heat capacity are denoted as bound points (separating positive/negative
temperature from each other). In particular, we will analyze the heat
capacity with fixed charge (in canonical ensemble) and with fixed chemical
potential (in grand canonical ensemble).

\subsubsection{Canonical ensemble}

Let us begin this subsection by computing the free energy of the system
which provides information regarding the amount of the work that a
thermodynamical system can perform. In total, this quantity is given by
removing the amount of energy that can not be used to perform work from
total internal energy of the system. The unusable energy is a combination of
the total entropy and temperature of system. Therefore, for these black
holes (in a canonical ensemble with a fixed charge $Q$), we have the
following Helmholtz free energy
\begin{equation}
F=M-TS=\frac{\Lambda \left( \varepsilon \right) r_{+}^{2}}{g\left(
\varepsilon \right) ^{2}f\left( \varepsilon \right) }-\frac{G\left(
\varepsilon \right) f\left( \varepsilon \right) q\left( \varepsilon \right)
^{2}}{4}\left[ \ln \left( \frac{r_{+}}{l(\varepsilon )}\right) -1\right] .
\label{free}
\end{equation}

The heat capacity with fixed charge is given by
\begin{equation}
C_{Q}=T\frac{\left( \frac{\partial S}{\partial r_{+}}\right) _{Q}}{\left(
\frac{\partial T}{\partial r_{+}}\right) _{Q}}=\frac{\pi r_{+}\left[
2\Lambda (\varepsilon )r_{+}^{2}-m(\varepsilon )^{2}c(\varepsilon
)c_{1}(\varepsilon )r_{+}+2f(\varepsilon )^{2}g(\varepsilon
)^{2}G(\varepsilon )q(\varepsilon )^{2}\right] }{4g(\varepsilon )\left[
\Lambda (\varepsilon )r_{+}^{2}-f(\varepsilon )^{2}g(\varepsilon
)^{2}G(\varepsilon )q(\varepsilon )^{2}\right] }.  \label{heat}
\end{equation}

From the above expression, the effects of gravity's rainbow on specific heat
can be seen easily.

\subsubsection{Grand canonical ensemble}

In the grand canonical ensemble with a fixed chemical potential (electric
potential, $U$, in this case) associated with the charge, the Gibbs free
energy for such black holes is given by
\begin{eqnarray}
\mathbb{G} &=&M-TS-\mu Q,  \nonumber \\
&=&\frac{\Lambda (\varepsilon )r_{+}^{2}}{4g(\varepsilon )^{2}}\left( \frac{1%
}{f(\varepsilon )}-\frac{1}{2}\right) +\frac{m(\varepsilon
)^{2}c(\varepsilon )c_{1}(\varepsilon )r_{+}}{8g(\varepsilon )^{2}}\left( 1-%
\frac{1}{f(\varepsilon )}\right)  \nonumber \\
&+&\frac{1}{4}\left( \frac{\mu (\varepsilon )}{\ln \left( \frac{r_{+}}{l}%
\right) }\right) ^{2}f(\varepsilon )\left[ G(\varepsilon )-\frac{%
f(\varepsilon )}{g(\varepsilon )^{2}}\ln \left( \frac{r_{+}}{l(\varepsilon )}%
\right) -2G(\varepsilon )\ln \left( \frac{r_{+}}{l(\varepsilon )}\right) %
\right] .
\end{eqnarray}

The Hawking temperature for the black holes in the grand canonical ensemble
is given by
\begin{equation}
T=-\frac{\Lambda \left( \varepsilon \right) r_{+}}{2\pi f\left( \varepsilon
\right) g\left( \varepsilon \right) }+\frac{m\left( \varepsilon \right)
^{2}c(\varepsilon )c_{1}(\varepsilon )}{4\pi f\left( \varepsilon \right)
g\left( \varepsilon \right) }-\frac{f\left( \varepsilon \right) g\left(
\varepsilon \right) G\left( \varepsilon \right) }{2\pi r_{+}}\left( \frac{%
\mu (\varepsilon )}{\ln \left( \frac{r_{+}}{l(\varepsilon )}\right) }\right)
^{2}.
\end{equation}

Now, the heat capacity with a fixed chemical potential is calculated by
\begin{equation}
C_{\mu }=T\left( \frac{\partial S}{\partial T}\right) _{\mu }=\frac{\pi
^{2}f(\varepsilon )g(\varepsilon )\left( \ln \left( \frac{r_{+}}{%
l(\varepsilon)}\right) \right) ^{3} \ r_{+}^{2} \ T}{f(\varepsilon
)^{2}g(\varepsilon )^{2}G(\varepsilon)\mu (\varepsilon )^{2}\left( \ln
\left( \frac{r_{+}}{l(\varepsilon )}\right) +2\right) -\Lambda (\varepsilon
)r_{+}^{2}\left( \ln \left( \frac{r_{+}}{l(\varepsilon )}\right) \right) ^{3}%
}.
\end{equation}

Next, we will study thermodynamical aspects of these black holes with the
help of obtained thermodynamical quantities.

\section{Thermodynamical aspects of charged massive BTZ black holes in
gravity's rainbow}

In this section, we are interested in studying, in particular, mass,
temperature and heat capacity of the charged massive BTZ black holes in
gravity's rainbow. We will discuss the free energy and phase diagram for
such black holes as well.

\subsection{Mass/Internal energy}

Our first item of the interest is mass of black holes. The total mass of
black holes has usually the interpretation of internal energy of a typical
system. Evidently, the mass has three distinctive terms: cosmological
constant term, $\Lambda(\varepsilon)$, massive term, $m(\varepsilon)$, and
charge term, $q(\varepsilon)$. Depending on the choices of different values
for these terms, the internal energy could have one of the following cases:

I) It is a positive definite function with a minimum. II) It is a positive
definite function everywhere except at the a point which is an extreme root.
III) It may have two roots with a region of negativity between these roots
and a minimum. IV) Being only an increasing function of the horizon radius
without any minimum.

Considering the positive nature of energy functions and other energy
dependent constants, we find that the charge term contributes to negativity
of the internal energy. As for $\Lambda(\varepsilon)$, its contribution
depends on the type of spacetime we are working in. For anti-de Sitter
spacetime, this term has constructive effects on the values of internal
energy. Whereas, for de Sitter case, the internal energy is a decreasing
function of this term. For the mass term, one finds that its effect depends
on the choices of $c_{1}(\varepsilon)$. For negative values of this
parameter, the mass term has negative effects on values of the internal
energy while the opposite effect is true for positive $c_{1}(\varepsilon)$.

In general, it is not possible to obtain the root of internal energy
analytically. But, regarding a vanishing term, it is possible to do so in
which the results are given as
\begin{eqnarray}
r_{1}|_{q(\varepsilon)=0} &=&{\frac{m( \varepsilon) ^{2}{c(\varepsilon )c}%
_{1}(\varepsilon )}{\Lambda ( \varepsilon) }},  \label{root of M chargeless}
\\
r_{2}|_{m(\varepsilon)=0} &=&l(\varepsilon )\ \exp \left[ -\frac{1}{2}\,%
\mathit{LambertW}\left( {\frac{\Lambda ( \varepsilon) {l}( \varepsilon) ^{2}%
}{G( \varepsilon) q( \varepsilon) ^{2}f( \varepsilon) ^{2}g( \varepsilon)
^{2}}}\right) \right],  \label{root of M massiveless} \\
r_{3}|_{\Lambda ( \varepsilon) =0} &=&\frac{-2\,G( \varepsilon) q(
\varepsilon) ^{2}f( \varepsilon) ^{2}g( \varepsilon) ^{2}\mathit{LambertW}%
\left( -{\frac{l( \varepsilon) m( \varepsilon) ^{2}{c(\varepsilon )c}%
_{1}(\varepsilon )}{2G( \varepsilon) q( \varepsilon) ^{2}f( \varepsilon)
^{2}g( \varepsilon) ^{2}}}\right) }{m( \varepsilon) ^{2}{c( \varepsilon) c}%
_{1}( \varepsilon) }.  \label{root of M Lambdaless}
\end{eqnarray}

As for the high energy limit of internal energy, the dominant term is the
charge term. In the absence of electric part of the solutions, for vanishing
horizon radius, hence evaporation, the total mass of black holes would
vanish too. Interestingly, it is possible to eliminate the effects of
electric part by setting, $l(\varepsilon )=r_{+} $. Therefore, for including
the effects of electric charge, the limit $l(\varepsilon )\neq r_{+}$ must
be satisfied. The second dominant term after the electric charge is the
massive term which highlights the effects of the massive gravity in high
energy regime. On the other hand, for asymptotical behavior, the leading
term will be the cosmological constant term. In the absence of this term,
the asymptotical behavior of the system will be governed by the massive term
which again, represents the effects of generalization to massive gravity.
Considering these two cases, one can conclude that for medium black holes,
the internal energy is highly affected by massive gravity. This means that
for medium black holes, the effects of the presence of massive gravitons
would be detectable within the internal energy. As for gravity's rainbow,
except for the root of mass in the absence of electric charge, the obtained
roots, high energy limit and asymptotical behavior are highly affected by
generalization to gravity's rainbow. Consequently, the extracted properties
and behaviors of the internal energy (existence of roots and negative values
for internal energy) depend on the choices of rainbow functions of the
metric, hence gravity's rainbow. Coupling different orders of rainbow
functions with different parameters provides the possibility of manipulation
of properties and behaviors of the internal energy.

\subsection{Temperature}

Now, let us focus on the temperature of these black holes. Here too, the
charge term has negative contribution on the values of temperature. The
effects of cosmological term depend on the spacetime under consideration.
For adS black holes, the cosmological term has positive effects on
temperature while for dS spacetime, the temperature is a decreasing function
of this term. Interestingly, the massive term is not coupled with any order
of horizon radius and it behaves as a constant. The roots of temperature are
marking bound points. The reason for such naming is as follows; in classical
thermodynamics, the negative values of temperature are interpreted as
non-physical solutions. Therefore, the roots of temperature separates
physical solutions form non-physical ones. It is a matter of calculation to
show that these black holes have following roots
\begin{equation}
r|_{T=0}=\,{\frac{m( \varepsilon) ^{2}\,c( \varepsilon) c_{1}( \varepsilon)
\pm \sqrt{m( \varepsilon) ^{4}\,c(\varepsilon )^{2}c_{1}(\varepsilon
)^{2}-16\,\Lambda ( \varepsilon) G( \varepsilon) q( \varepsilon) ^{2}f(
\varepsilon) ^{2}g( \varepsilon) ^{2}}}{4\Lambda ( \varepsilon) }}.
\label{root of
Temperature}
\end{equation}%
The existence of real valued root for the temperature is limited to
following condition
\begin{equation}
m( \varepsilon) ^{4}\,c(\varepsilon )^{2}c_{1}(\varepsilon )^{2}-16\,\Lambda
( \varepsilon) G( \varepsilon) q( \varepsilon) ^{2}f( \varepsilon) ^{2}g(
\varepsilon) ^{2}\geq 0,  \label{condition1}
\end{equation}%
which could be used to extract a specific limitation for the mass of
graviton in term of other parameters
\begin{equation}
m( \varepsilon) \,=\left( {\frac{16\,\Lambda ( \varepsilon) G( \varepsilon)
q( \varepsilon) ^{2}f( \varepsilon) ^{2}g( \varepsilon) ^{2}}{c(\varepsilon
)^{2}c_{1}(\varepsilon )^{2}}} \right)^{\frac{1}{4}}.
\label{mass of graviton}
\end{equation}

Once more, we emphasize that by satisfying obtained condition, the roots of
temperature will be real valued. For adS black holes, only one positive
valued root exists for the temperature which is the negative branch of the
obtained roots. On the contrary, for dS black holes, two positive valued
roots may exist for the temperature. The existence of second positive valued
root depends on the following condition
\begin{equation}
0<\sqrt{m( \varepsilon) ^{4}\,c(\varepsilon )^{2}c_{1}(\varepsilon
)^{2}-16\,\Lambda ( \varepsilon) G( \varepsilon) q( \varepsilon) ^{2}f(
\varepsilon) ^{2}g( \varepsilon) ^{2}}<m( \varepsilon) ^{2}\,c(\varepsilon
)c_{1}(\varepsilon ),  \label{condition2}
\end{equation}%
which is partly similar to the condition for having real valued roots. The
resulting roots show that the contributions of rainbow functions of the
metric, could only be observed in the electric charge term. It means that in
the roots of temperature, the energy functions of metric are only coupled
with the electric charge term.

In the absence of electric charge, the root of temperature is given by
\begin{equation}
r_{T=0\text{ }with\text{ }q(\varepsilon )=0}={\frac{m(\varepsilon
)^{2}\,c(\varepsilon )c_{1}(\varepsilon )}{2\Lambda (\varepsilon )}},
\label{root of Temperature chargeless}
\end{equation}%
which shows that the positive valued root exists only for dS spacetime and
this root is an increasing function of the graviton's mass and a decreasing
function of the cosmological constant. Interestingly, in this case, the
rainbow functions have no effects on the root of temperature.

For massless gravitons case, the root of temperature will be obtained as
\begin{equation}
r_{T=0\text{ }with\text{ }m(\varepsilon) =0}=-{\frac{\sqrt{-\Lambda (
\varepsilon) G( \varepsilon) }g( \varepsilon) f( \varepsilon) q(
\varepsilon) }{\Lambda ( \varepsilon) }}.
\label{root of Temperature massiveless}
\end{equation}

Evidently, for this case, the real valued root only exists for adS black
holes and only one positive root could be extracted for these black holes in
this case. Here, the root is an increasing function of the electric charge
and rainbow functions.

For the absence of cosmological constant, the root of temperature could be
calculated as
\begin{equation}
r_{T=0\text{ }with\text{ }\Lambda \left( \varepsilon\right) =0}=2\,{\frac{G(
\varepsilon) q( \varepsilon) ^{2}f( \varepsilon) ^{2}g( \varepsilon) ^{2}}{%
m( \varepsilon) ^{2}\,c(\varepsilon )c_{1}(\varepsilon )}},
\label{root of Temperature Lambdaless}
\end{equation}%
which, contrary to the absence of electric charge case, is a decreasing
function of the graviton's mass. It is worthwhile to mention that in this
case, the root is an increasing function of the electric charge and rainbow
functions.

The high temperature limit of these black holes are governed by the electric
charge term. On the other hand, the leading order in asymptotical behavior
is the cosmological constant term. Interestingly, in the absence of electric
charge (whether setting electric charge zero or consider the case $%
l(\varepsilon)=r_{+}$), the next leading order in high temperature will be
the massive term. Now, remembering that this term in the temperature is not
coupled with any order of the horizon radius, one can see that for the
evaporation of black holes (vanishing horizon radius) the temperature will
be non-zero. In other words, in the evaporation of these black holes, a
trace of the existence of black holes is left behind which presents itself
as fluctuation in the temperature of spacetime. This provides the
possibility of the existence of black hole' information after their
evaporation. This remnant, for the temperature of black holes, is an
increasing function of the graviton's mass and a decreasing function of the
rainbow functions. Considering the effects of massive gravity, one can state
that thermodynamical behavior of temperature for medium black holes is
governed by the mass of graviton. On the other hand, the effects of
gravity's rainbow on the temperature could be observed for the three cases
of small, medium and large black holes. In other words, the generalization
of gravity's rainbow, contrary to massive gravity, has some effects on the
temperature of all black holes with different sizes. It is worthwhile to
mention that the effects of gravity's rainbow on the temperature of medium
and large black holes are the same while for the small black holes, these
effects are opposite.

\subsection{Heat capacity}

By taking a closer look at the heat capacity, one can see that its numerator
is the same as temperature. Therefore, the roots of heat capacity and
temperature are the same, and the arguments that were stated in the last
subsection (for the temperature and its roots) can apply for the heat
capacity and its roots as well. On the other hand, the denominator of heat
capacity contains information regarding phase transition points. In other
words, the divergence point of heat capacity (roots of denominator of the
heat capacity) can be characterized as a phase transition.

Denominator of the heat capacity contains only electric charge and
cosmological constant terms with coupling of rainbow functions and so it
does not depend on the massive gravity parameter. In other words, the
generalization to gravity's rainbow affects the divergencies of heat
capacity, hence phase transitions of the black holes, whereas the existence
of massive gravitons does not affect the phase transitions of these black
holes. It is a matter of calculation to show that by using Eq. (\ref{heat}),
the positive valued divergence point of heat capacity is obtained as
\begin{equation}
r_{_{C_{Q} \rightarrow \infty}}={\frac{\sqrt{\Lambda ( \varepsilon) G(
\varepsilon) }g( \varepsilon) f( \varepsilon) q( \varepsilon) }{\Lambda (
\varepsilon) }}.  \label{divergency of
heat}
\end{equation}

Evidently, the real valued phase transition points are observed for dS black
holes only. The existence of the phase transition point depends on the
electric charge and cosmological terms. This phase transition point is an
increasing function of the electric charge and rainbow functions.

The high energy limit of the heat capacity is given by
\begin{equation}
\lim_{very\text{ }small\text{ }r_{+}}C_{Q}=-\,{\frac{\pi }{2g( \varepsilon) }%
}\,r_{+}+\,{\frac{\pi m( \varepsilon) ^{2}\,c(\varepsilon )c_{1}(\varepsilon
)}{4g( \varepsilon) ^{3}G( \varepsilon) q( \varepsilon) ^{2}f( \varepsilon)
^{2}}}{r}_{+}^{2}-{\frac{\Lambda ( \varepsilon) \pi }{g( \varepsilon) ^{3}G(
\varepsilon) q( \varepsilon) ^{2}f( \varepsilon) ^{2}}}r_{+}^{3}+O\left(
r_{+}^{4}\right) ,  \label{high energy limit heat}
\end{equation}%
in which the effects of generalization to gravity's rainbow could be
observed in dominant term. The presence of electric charge and massive
gravity could be detected from second dominant term of the high energy
limit. Here, the massive parameter and electric charge are coupled with each
other. The contribution of the cosmological constant could be seen in third
leading order, which is coupled with the electric charge as well. Taking a
closer look, one can see that the effects of gravity's rainbow are presented
in all three leading orders of high energy limit of the heat capacity. As
one can see, the effects of massive gravity could be more highlighted for
the medium black holes while for large black holes, the effect of
cosmological constant governs the behavior of heat capacity.

Interestingly, in the absence of electric charge (similar to the cases that
were studied in temperature), the dominant term for the high energy limit
behaves like a constant including massive gravity and cosmological constant
in following form
\begin{equation}
\lim_{very\text{ }small\text{ }r_{+}}C_{Q}=-\,{\frac{\pi m( \varepsilon)
^{2}\,c(\varepsilon )c_{1}(\varepsilon )}{4g( \varepsilon) \Lambda (
\varepsilon) }}+\,{\frac{\pi }{2g( \varepsilon) }}\,r_{+}+O\left(
r_{+}^{2}\right), \text{ \ \ \ \ \ }\mbox{for} \text{ \ \ \ \ }q(
\varepsilon) =0.  \label{high energy
limit heat chargeless}
\end{equation}

Similar to the temperature, here too, for vanishing horizon radius
(evaporation of these black holes), the heat capacity will be non-zero. This
shows that the traces of the existence of black holes after their
evaporation could be observed in differences of heat capacity of the place
where black holes existed. In this case, one can see that graviton's mass
modify thermodynamical behavior of the black holes in their last stage of
existence. The modifications of the gravity's rainbow could be seen by the
presence of rainbow function, $g(\varepsilon)$.

On the other hand, for the asymptotic limit, one can derive following
relation for the heat capacity
\begin{equation}
\lim_{very\text{ \textit{large} }r_{+}}C_{Q}=\,{\frac{\pi }{2g( \varepsilon)
}}r_{+}-{\frac{\pi m( \varepsilon) ^{2}\,c(\varepsilon )c_{1}(\varepsilon )}{%
4g( \varepsilon) \Lambda ( \varepsilon) }}+{\frac{G( \varepsilon) q(
\varepsilon) ^{2}f( \varepsilon) ^{2}g( \varepsilon) \pi }{\Lambda (
\varepsilon) r_{+}}}+O\left( r_{+}^{-2}\right) .  \label{asymptotic heat}
\end{equation}

First of all, the dominant term for the asymptotical behavior of heat
capacity is same as one extracted for the high energy limit with different
sign. The second leading term in this case behaves like a constant term
including massive gravity which, opposite to high energy limit, is coupled
with the cosmological constant. The presence of electric charge part of the
solutions could be observed in the third leading term of this case. Like
previous, here, the presence of gravity's rainbow could be observed in all
the three leading terms in asymptotical behavior of the heat capacity. One
of the differences of this case with the high energy limit is the fact that
in this limit, the presence of electric charge (cosmological constant) was
only observed in denominator (numerator) of leading terms whereas, in the
asymptotical case, the presence of electric charge (cosmological constant)
was only observed in numerator (denominator) of the leading terms. In the
absence of cosmological constant, the asymptotical behavior of heat capacity
will be modified into
\begin{equation}
\lim_{very\text{ \textit{large} }r_{+}}C_{Q}=\,{\frac{\pi m( \varepsilon)
^{2}\,c(\varepsilon )c_{1}(\varepsilon )}{4g( \varepsilon) ^{3}G(
\varepsilon) q( \varepsilon) ^{2}f( \varepsilon) ^{2}}}r_{+}^2-\,{\frac{\pi
}{2g( \varepsilon) }}r_{+}\text{\ \ \ \ }for\text{ }\Lambda ( \varepsilon)
=0,  \label{asymptotic Lambdaless}
\end{equation}%
which shows that the dominant term in the asymptotical behavior of heat
capacity is a coupling between massive gravity and electric part of the
solutions with the effects of gravity's rainbow. Considering the effects of
massive gravity in both high energy regime and asymptotic limit, one can see
that graviton's mass highly modifies the behavior of heat capacity in both
of these regimes. This highlights the contribution of massive gravity in
thermodynamical behavior of these black holes. The same could be also stated
for gravity's rainbow due to the presence of rainbow functions in both of
these regimes in the leading terms.

\subsubsection{Free energy}

The free energy contains information regarding the phase transition points.
The chemical equilibrium is reached for a system when its free energy is
minimized. In other words, the first order derivation of the free energy
with respect to thermodynamical quantities vanishes at the equilibrium
point. This equilibrium point marks the place where system goes under a
phase transition. The other important information regarding the free energy
is stored in its roots. In general, the root of free energy for these black
holes is
\begin{equation}
r_{_{F=0}}=l(\varepsilon ) \ \exp \left[
-\frac{1}{2}\mathit{LambertW}\left( -{\frac{8\Lambda (
\varepsilon) {l(\varepsilon)}^{2}{\mathrm{\exp }}\left( 2\right)
}{G( \varepsilon) q( \varepsilon) ^{2}f( \varepsilon) ^{2}g(
\varepsilon) ^{2}}}\right) +1\right] .  \label{root of free}
\end{equation}

Using the concept which was introduced for obtaining phase transition point
from the free energy, one can show that the positive valued extremum point
of the free energy is obtained as
\begin{equation}
r_{_{F \rightarrow \infty}}={\frac{\sqrt{\Lambda ( \varepsilon) G(
\varepsilon) }g( \varepsilon) f( \varepsilon) q( \varepsilon) }{\Lambda (
\varepsilon) }},
\end{equation}%
which is exactly the same phase transition point obtained for the heat
capacity. Therefore, divergencies of the heat capacity and extremum of the
free energy coincide with each other. The high energy limit of free energy
is governed by the electric charge term. But here, similar to the case of
internal energy, it is possible to cancel the effects of charge term through
setting $l(\varepsilon)=r_{+}(\varepsilon)$. On the contrary, the leading
order in the asymptotical behavior of free energy is the cosmological
constant. In both of the mentioned regimes, the effects of gravity's rainbow
could be observed by the coupling of energy functions with different
parameters. It is worthwhile to mention that the free energy is independent
of the generalization to massive gravity. In other words, the free energy
(energy which could be converted to work) is independent of the graviton's
mass.

Using the obtained extremum (critical horizon radius), one can obtain
internal energy, temperature and free energy of the phase transition point
as
\begin{eqnarray}
M_{_{Phase\text{ }Transition}} &=&\frac{\,q( \varepsilon) }{8g( \varepsilon)
\Lambda ( \varepsilon) }\left( m( \varepsilon) ^{2}c(\varepsilon
)c_{1}(\varepsilon )\sqrt{\Lambda ( \varepsilon) G( \varepsilon) }-f(
\varepsilon) G( \varepsilon) q( \varepsilon) g( \varepsilon) \Lambda (
\varepsilon) \right.  \nonumber \\
&&\left. -\left[ 1+2\,\ln \left( {\frac{\sqrt{\Lambda ( \varepsilon) G(
\varepsilon) }g( \varepsilon) f( \varepsilon) q( \varepsilon) }{\Lambda (
\varepsilon) l(\varepsilon )}}\right) \right] \right) ,  \label{Phase mass}
\\
T_{_{Phase\text{ }Transition}} &=&\,{\frac{m( \varepsilon) ^{2}c(\varepsilon
)c_{1}(\varepsilon )\sqrt{\Lambda ( \varepsilon) G( \varepsilon) }-4\,f(
\varepsilon) G( \varepsilon) q( \varepsilon) g( \varepsilon) \Lambda (
\varepsilon) }{4\pi \,f( \varepsilon) g( \varepsilon) \sqrt{\Lambda (
\varepsilon) G( \varepsilon) }}},  \label{phase temp} \\
F_{_{Phase\text{ }Transition}} &=&\frac{\,f( \varepsilon) G( \varepsilon) q(
\varepsilon) ^{2}}{8}\left[ 3-2\,\ln \left( {\frac{\sqrt{\Lambda (
\varepsilon) G( \varepsilon) }g( \varepsilon) f( \varepsilon) q(
\varepsilon) }{\Lambda ( \varepsilon) l(\varepsilon )}}\right) \right],
\label{phase free}
\end{eqnarray}
where we should regard the positive cosmological constant to obtain real
valued quantities.

\subsubsection{Phase diagrams}

\begin{figure}[tbp]
$%
\begin{array}{ccc}
\epsfxsize=5cm \epsffile{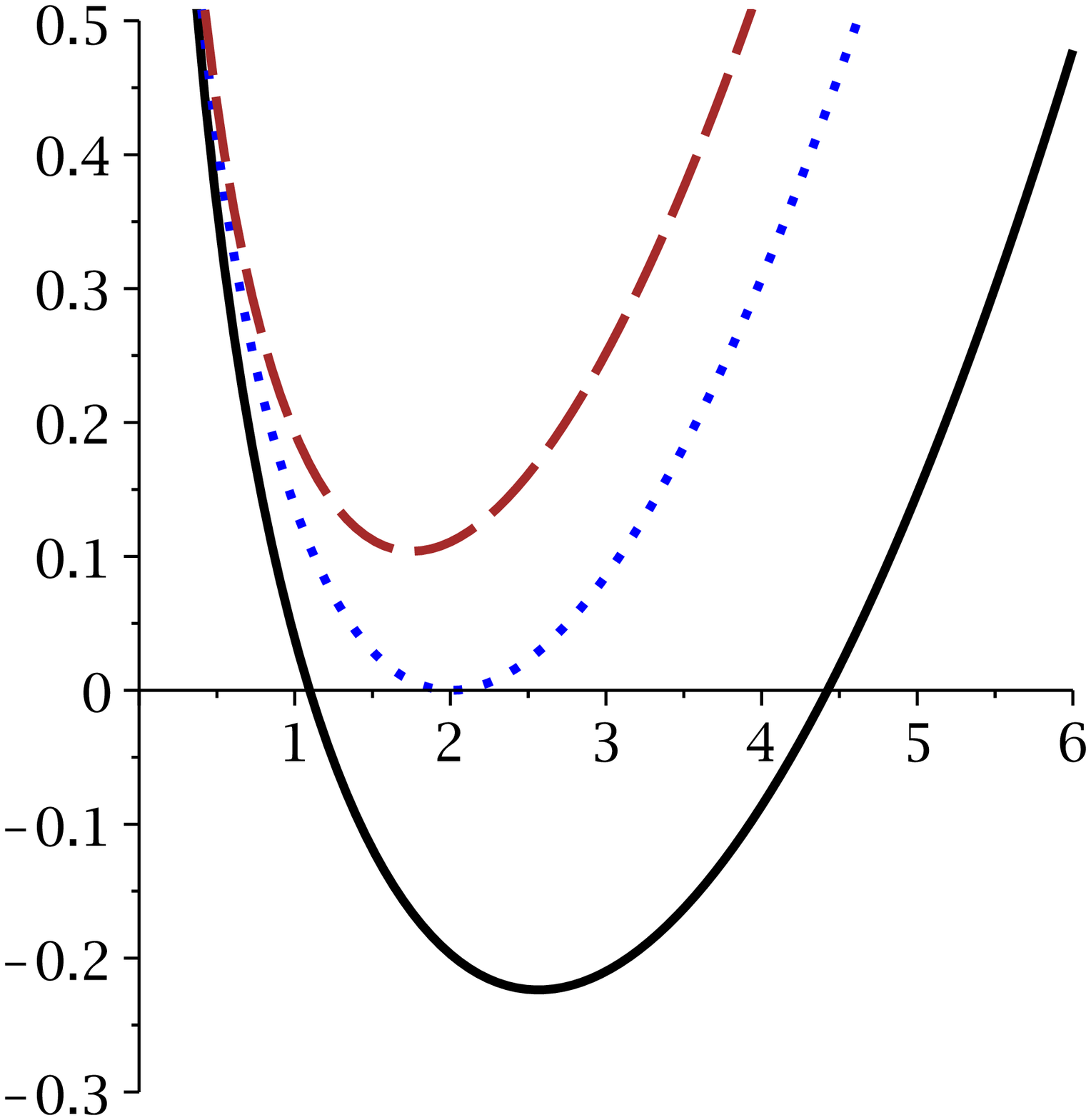} & \epsfxsize=5cm %
\epsffile{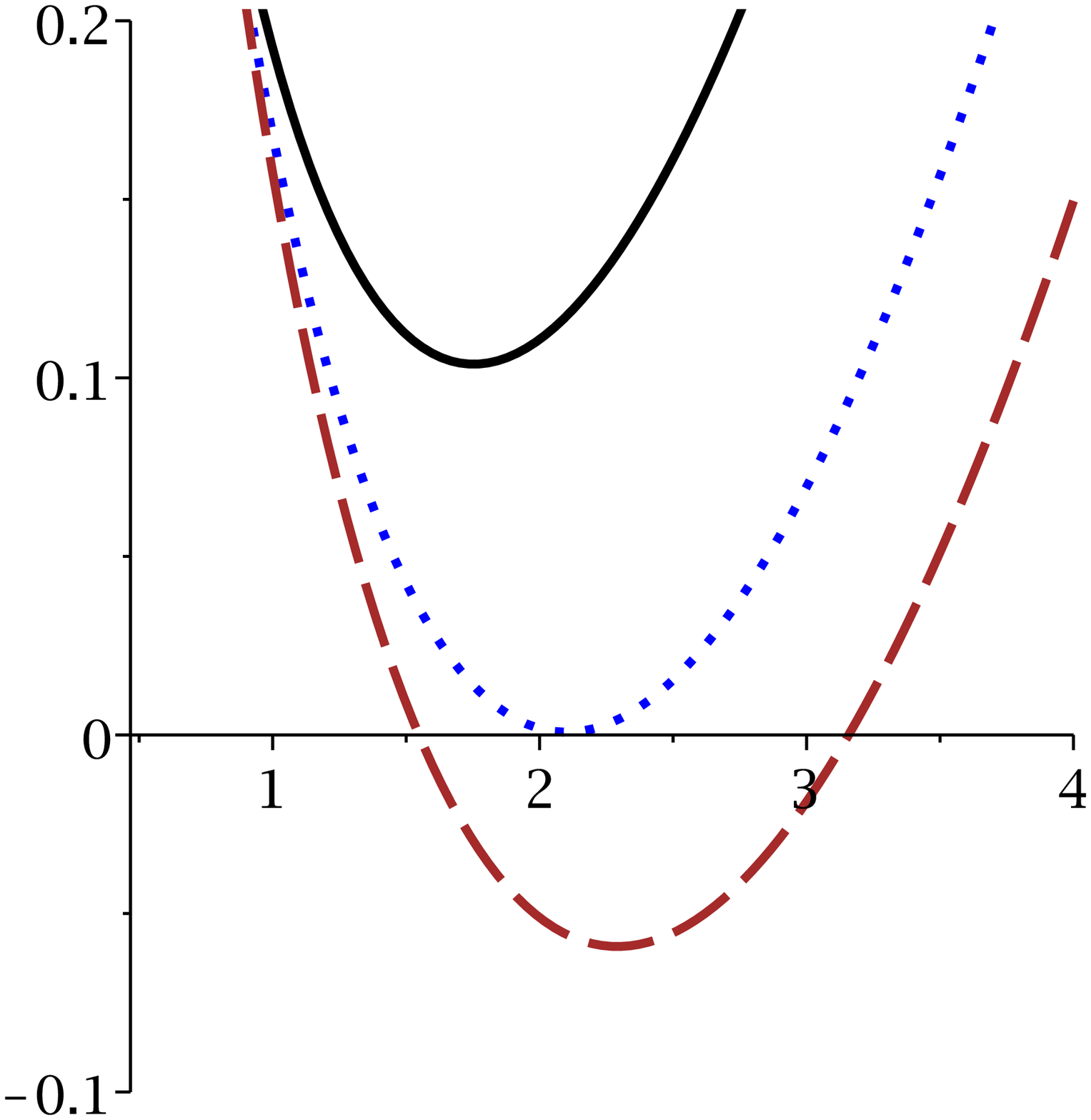} & \epsfxsize=5cm \epsffile{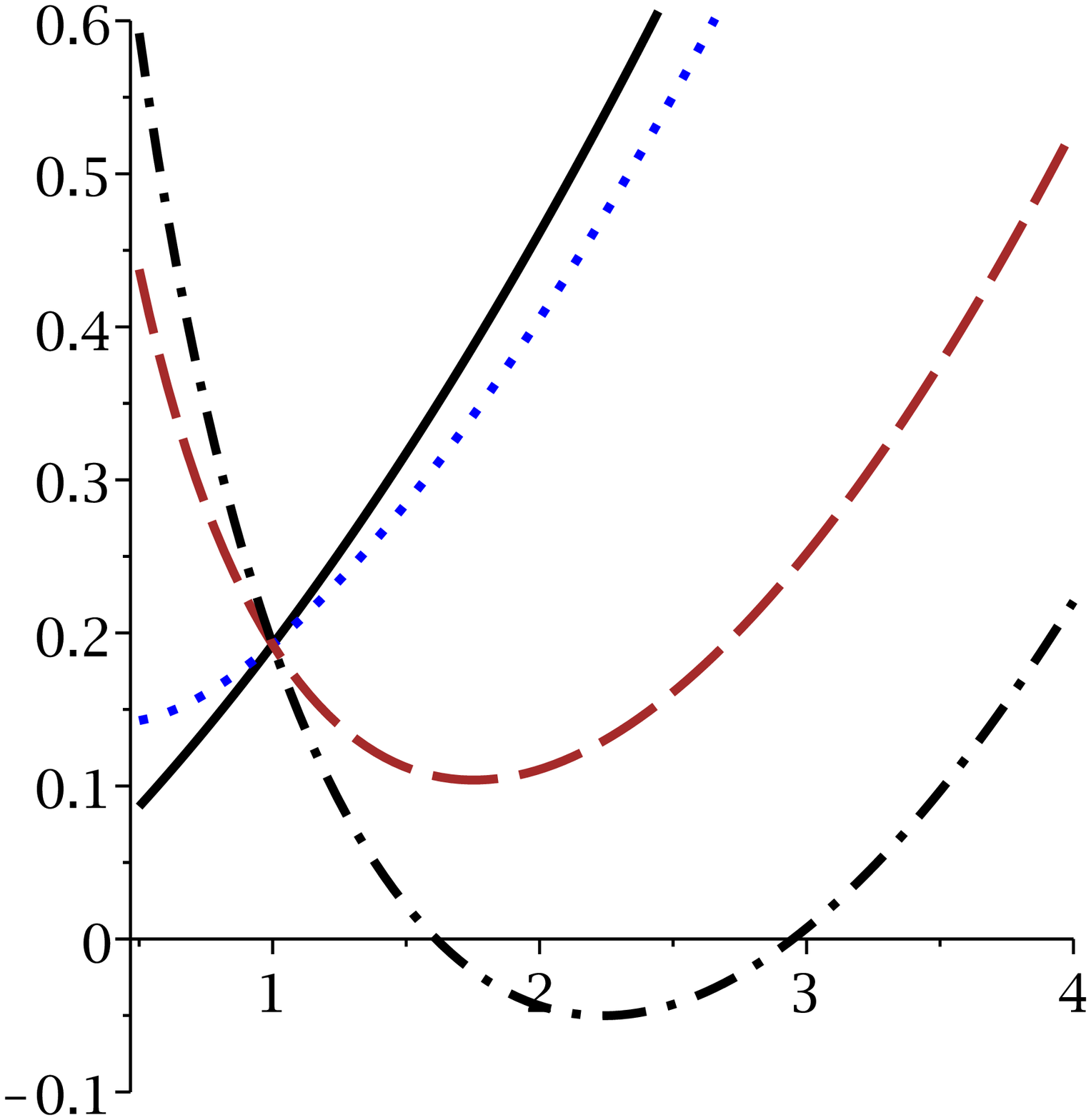}%
\end{array}
$%
\caption{For different scales: $M$ versus $r_{+}$ for $G(\protect\varepsilon%
)=l (\protect\varepsilon)=1$, $c(\protect\varepsilon)=c_{1}(\protect%
\varepsilon)=2$, $g(\protect\varepsilon)=1.9$ and $\Lambda(\protect%
\varepsilon)=-1$. \newline
Left panel: $q(\protect\varepsilon)=1.5$, $f(\protect\varepsilon)=0.9$, $m(%
\protect\varepsilon)=0$ (continuous line), $m(\protect\varepsilon)=0.8$
(dotted line) and $m(\protect\varepsilon)=1$ (dashed line). \newline
Middle panel: $q(\protect\varepsilon)=1.5$, $m(\protect\varepsilon)=1$, $f(%
\protect\varepsilon)=0.9$ (continuous line), $f(\protect\varepsilon)=1.03$
(dotted line) and $f(\protect\varepsilon)=1.1$ (dashed line). \newline
Right panel: $f(\protect\varepsilon)=0.9$, $m(\protect\varepsilon)=1$, $q(%
\protect\varepsilon)=0$ (continuous line), $q(\protect\varepsilon)=0.6$
(dotted line), $q(\protect\varepsilon)=1.5$ (dashed line) and $q(\protect%
\varepsilon)=1.8$ (dashed-dotted line).}
\label{Fig3}
\end{figure}

\begin{figure}[tbp]
$%
\begin{array}{ccc}
\epsfxsize=5cm \epsffile{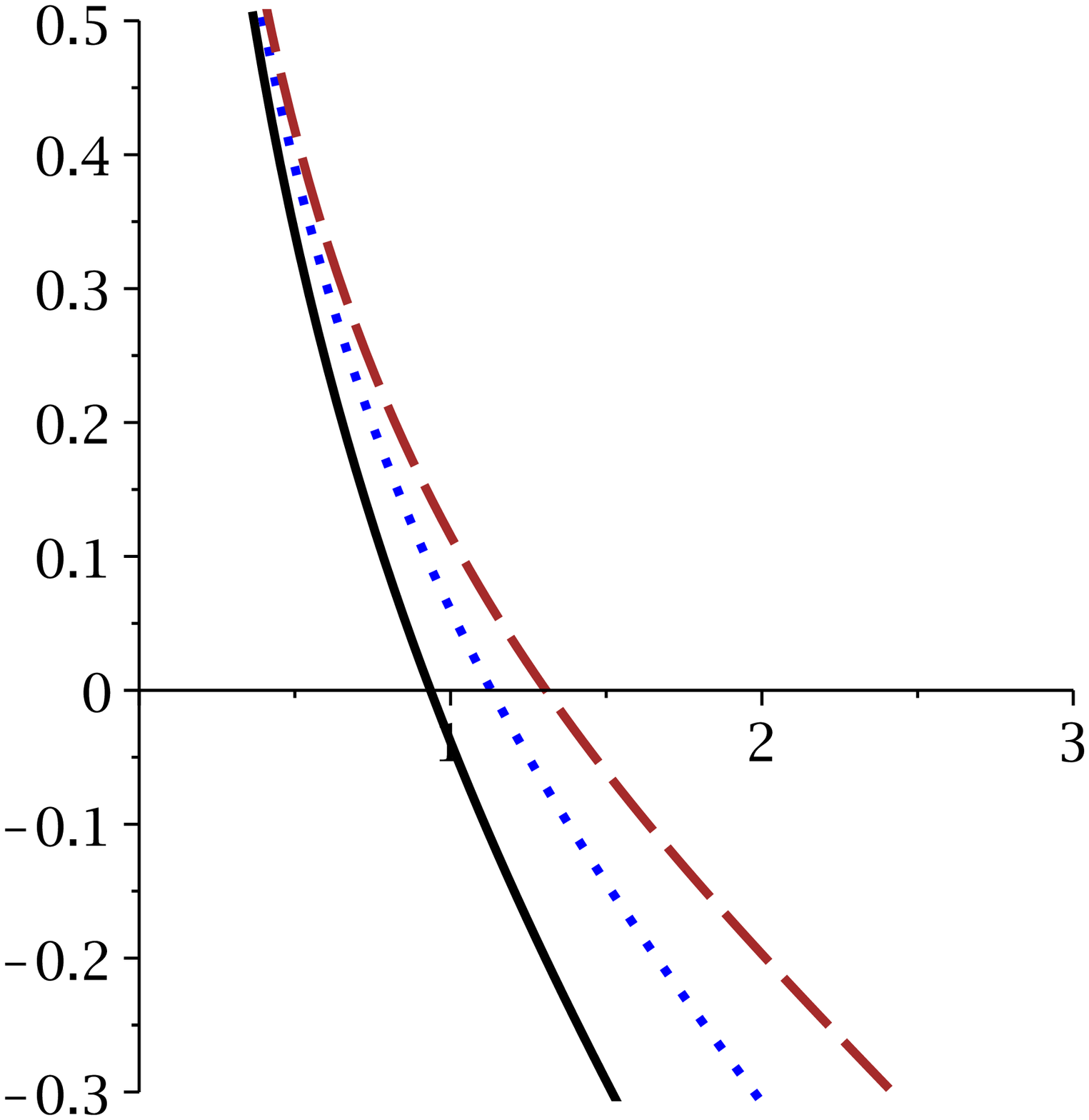} & \epsfxsize=5cm %
\epsffile{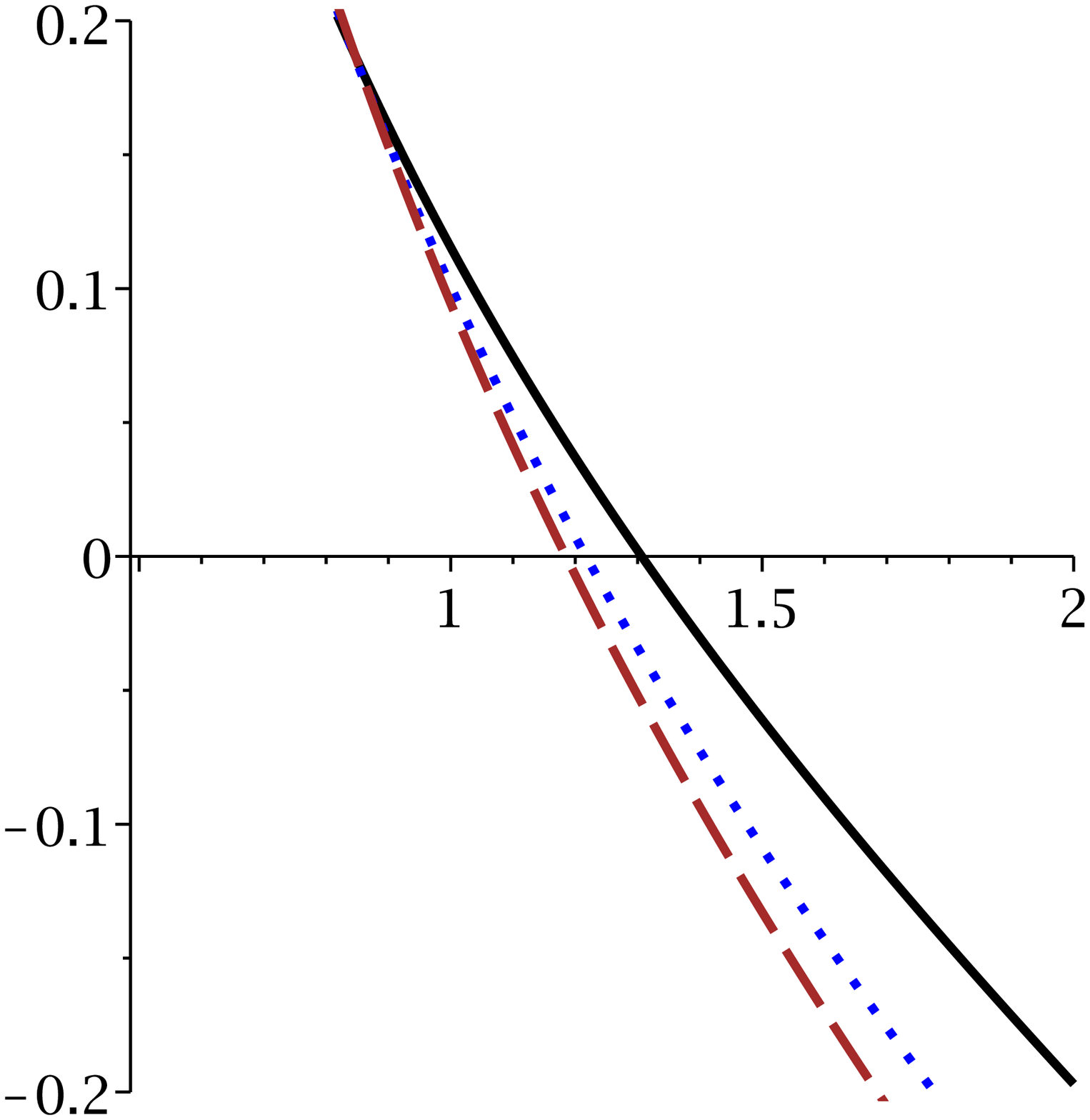} & \epsfxsize=5cm \epsffile{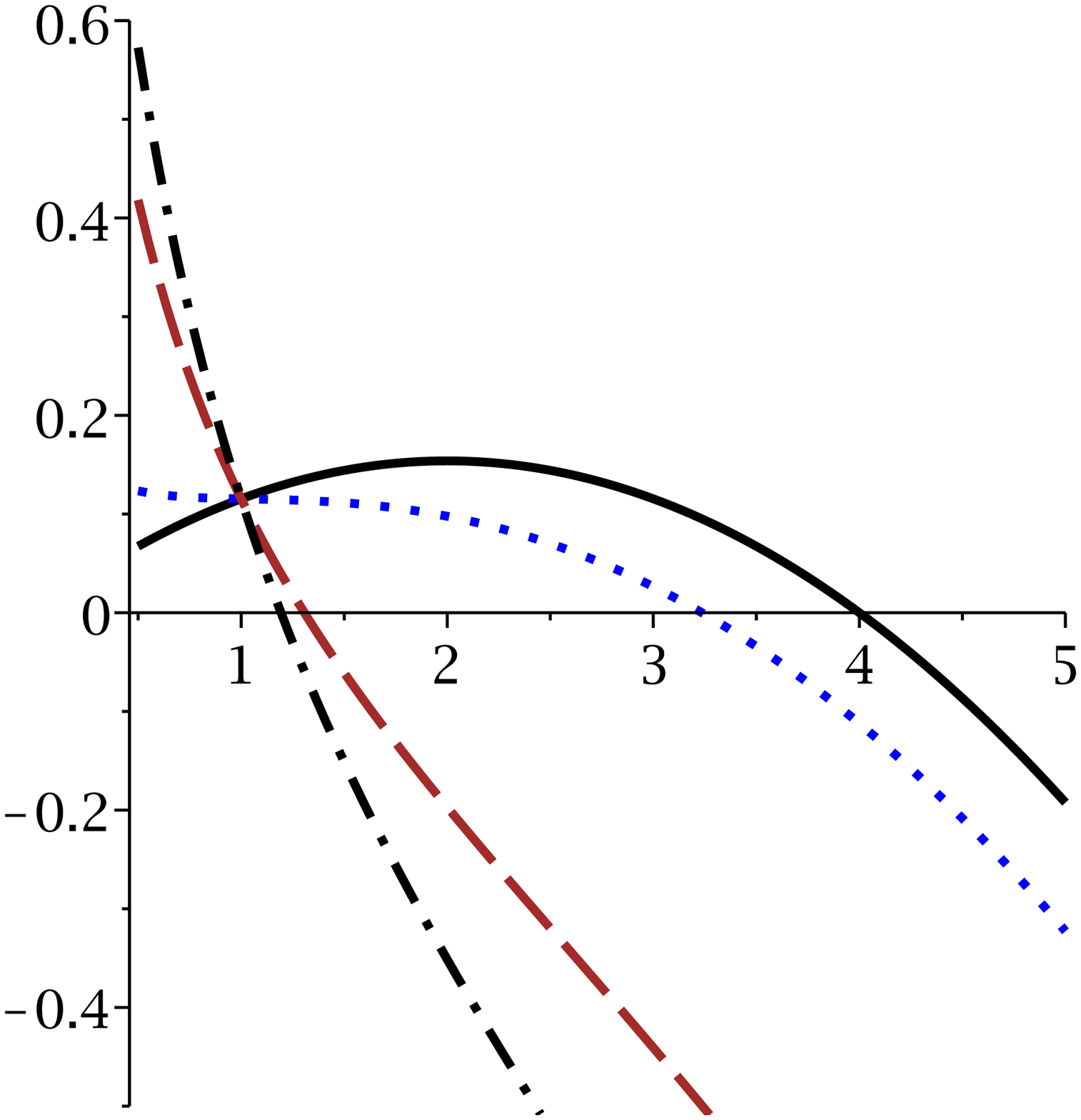}%
\end{array}
$%
\caption{For different scales: $M$ versus $r_{+}$ for $G(\protect\varepsilon%
)=l(\protect\varepsilon)=1$, $c(\protect\varepsilon)=c_{1}(\protect%
\varepsilon)=2$, $g(\protect\varepsilon)=1.9$ and $\Lambda(\protect%
\varepsilon)=1$. \newline
Left panel: $q(\protect\varepsilon)=1.5$, $f(\protect\varepsilon)=0.9$, $m(%
\protect\varepsilon)=0$ (continuous line), $m(\protect\varepsilon)=0.8$
(dotted line) and $m(\protect\varepsilon)=1$ (dashed line). \newline
Middle panel: $q(\protect\varepsilon)=1.5$, $m(\protect\varepsilon)=1$, $f(%
\protect\varepsilon)=0.9$ (continuous line), $f(\protect\varepsilon)=1.03$
(dotted line) and $f(\protect\varepsilon)=1.1$ (dashed line). \newline
Right panel: $f(\protect\varepsilon)=0.9$, $m(\protect\varepsilon)=1$, $q(%
\protect\varepsilon)=0$ (continuous line), $q(\protect\varepsilon)=0.6$
(dotted line), $q(\protect\varepsilon)=1.5$ (dashed line) and $q(\protect%
\varepsilon)=1.8$ (dashed-dotted line).}
\label{Fig4}
\end{figure}

\begin{figure}[tbp]
$%
\begin{array}{ccc}
\epsfxsize=5cm \epsffile{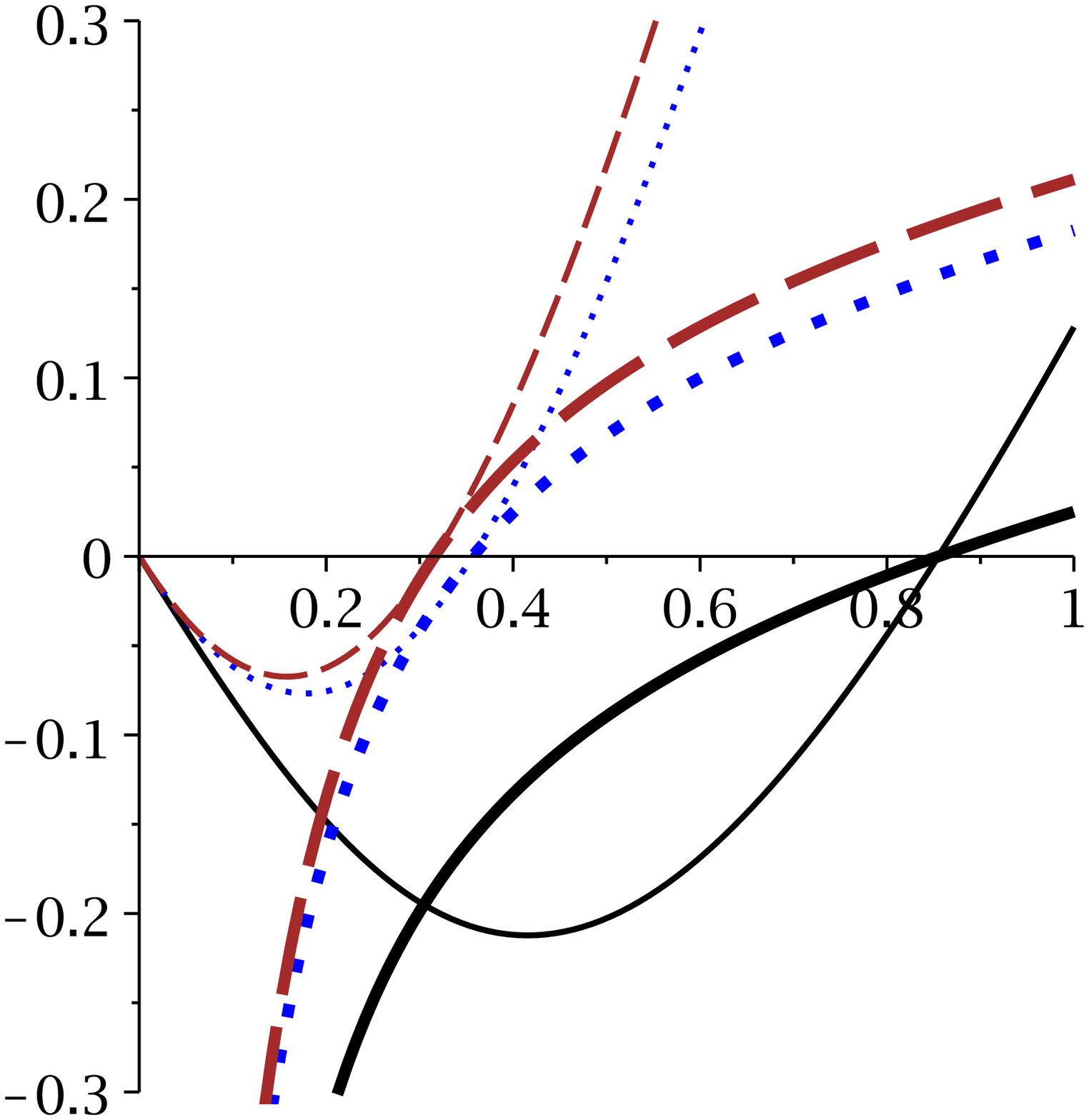} & \epsfxsize=5cm %
\epsffile{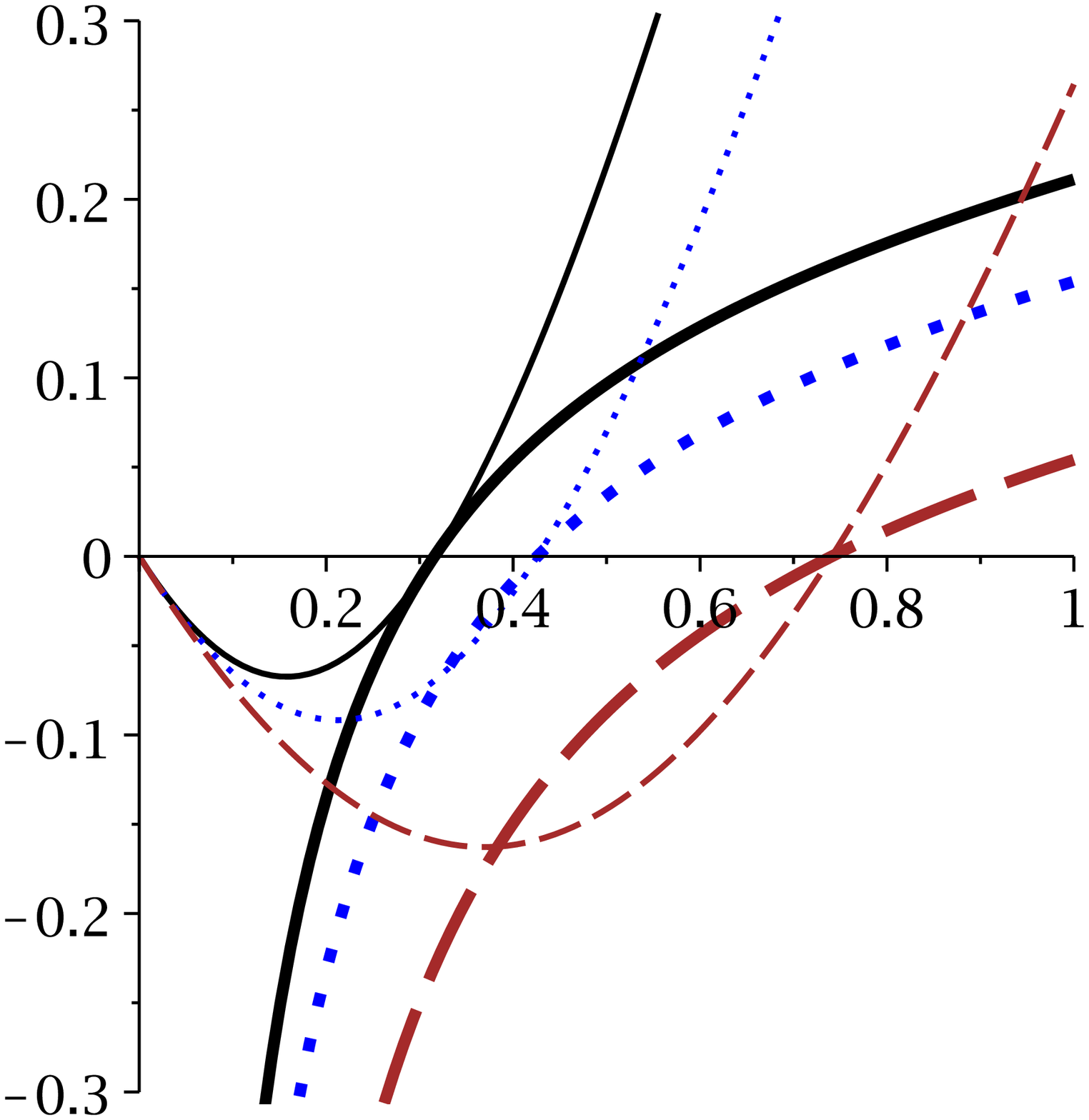} & \epsfxsize=5cm %
\epsffile{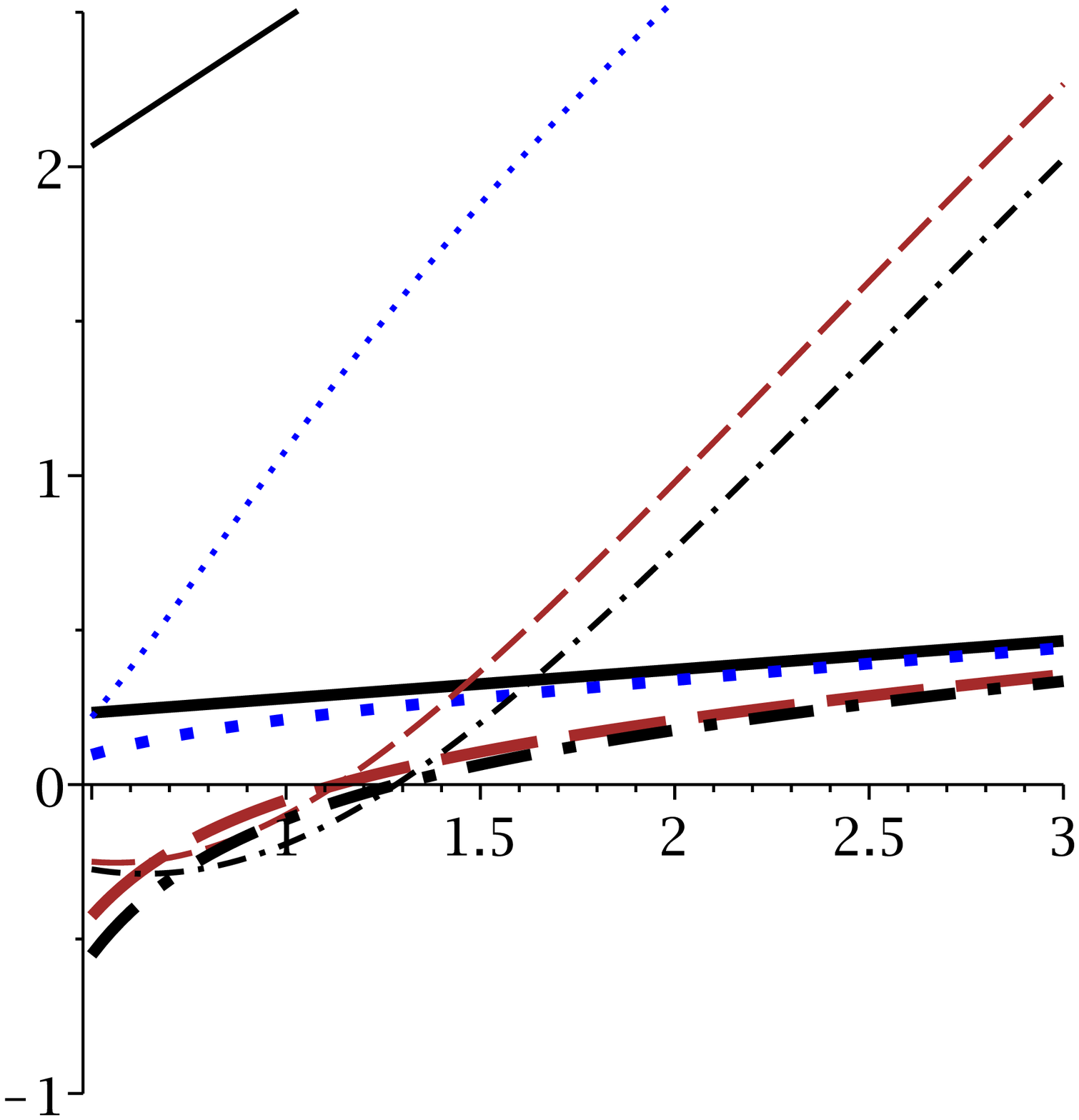}%
\end{array}
$%
\caption{For different scales: $C_{Q}$ and $T$ (bold lines) versus $r_{+}$
for $G(\protect\varepsilon)=l(\protect\varepsilon)=1$, $c(\protect\varepsilon%
)=c_{1}(\protect\varepsilon)=2$, $g(\protect\varepsilon)=1.9$ and $\Lambda(%
\protect\varepsilon)=-1$. \newline
Left panel: $q(\protect\varepsilon)=0.5$, $f(\protect\varepsilon)=0.9$, $m(%
\protect\varepsilon)=0$ (continuous line), $m(\protect\varepsilon)=0.92$
(dotted line) and $m(\protect\varepsilon)=1$ (dashed line). \newline
Middle panel: $q(\protect\varepsilon)=0.5$, $m(\protect\varepsilon)=1$, $f(%
\protect\varepsilon)=0.9$ (continuous line), $f(\protect\varepsilon)=1.07$
(dotted line) and $f(\protect\varepsilon)=1.5$ (dashed line). \newline
Right panel: $f(\protect\varepsilon)=0.9$, $m(\protect\varepsilon)=1$, $q(%
\protect\varepsilon)=0$ (continuous line), $q(\protect\varepsilon)=0.5$
(dotted line), $q(\protect\varepsilon)=1.1$ (dashed line) and $q(\protect%
\varepsilon)=1.2$ (dashed-dotted line).}
\label{Fig5}
\end{figure}

\begin{figure}[tbp]
$%
\begin{array}{ccc}
\epsfxsize=5cm \epsffile{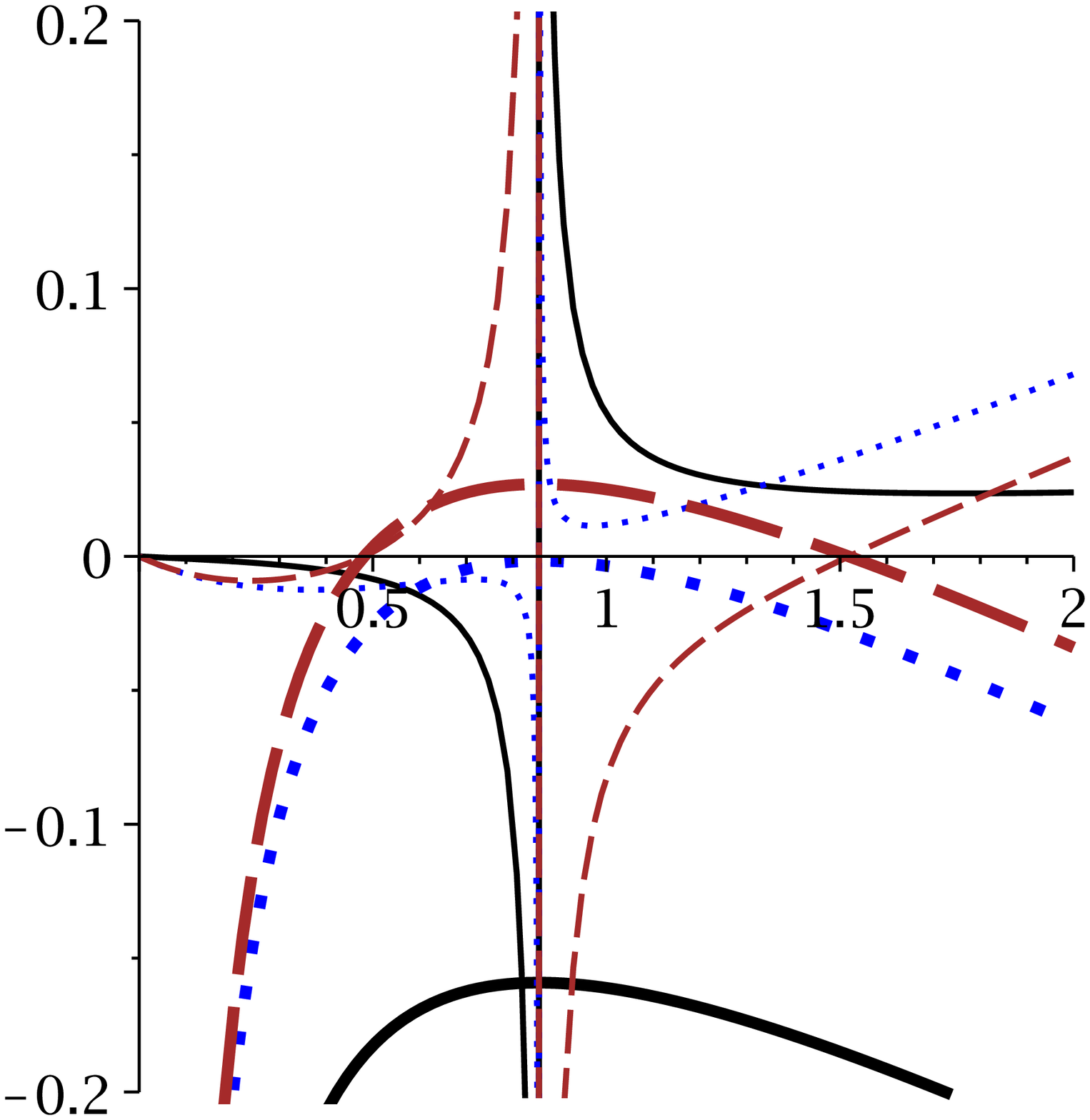} & \epsfxsize=5cm %
\epsffile{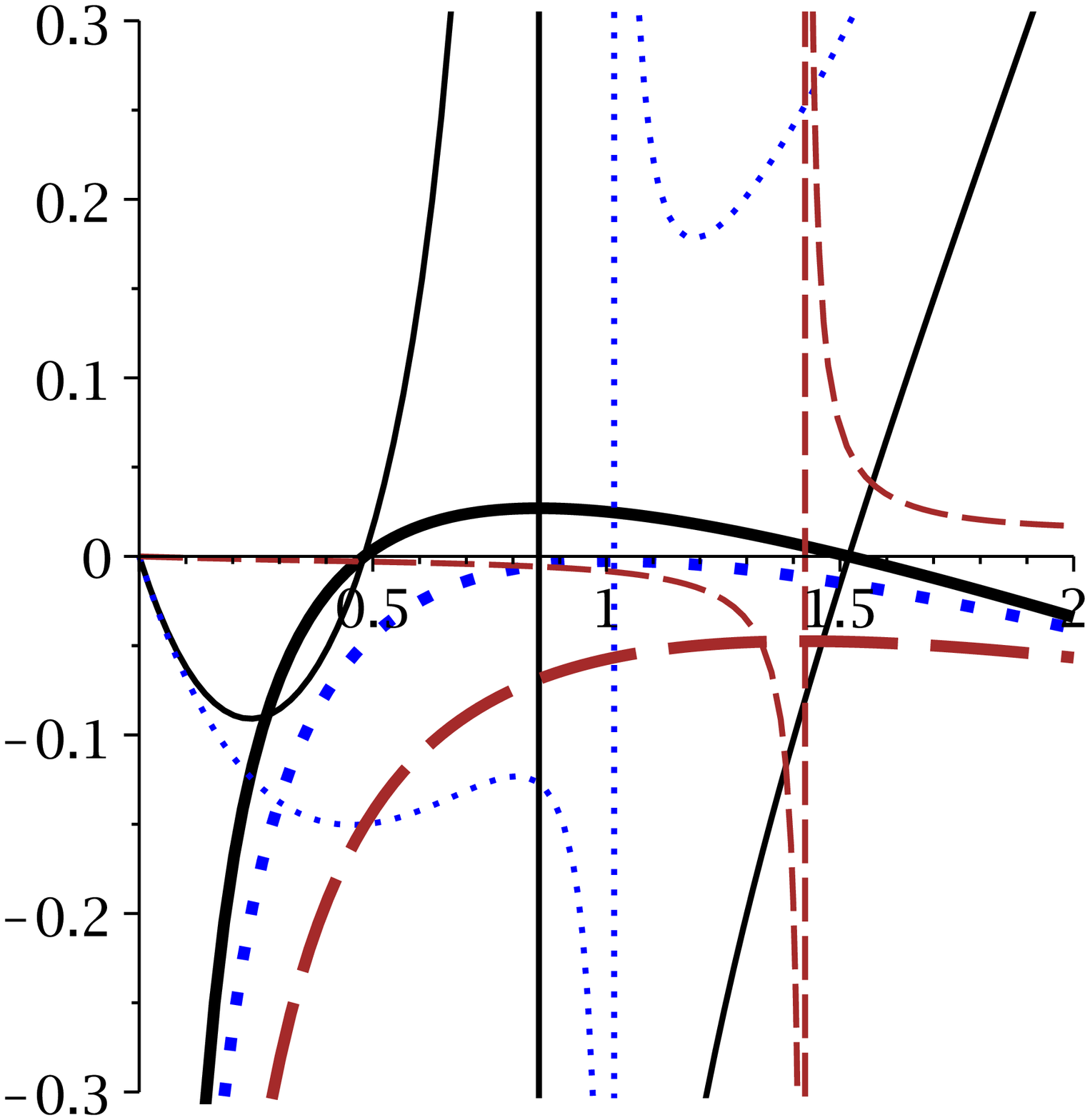} & \epsfxsize=5cm \epsffile{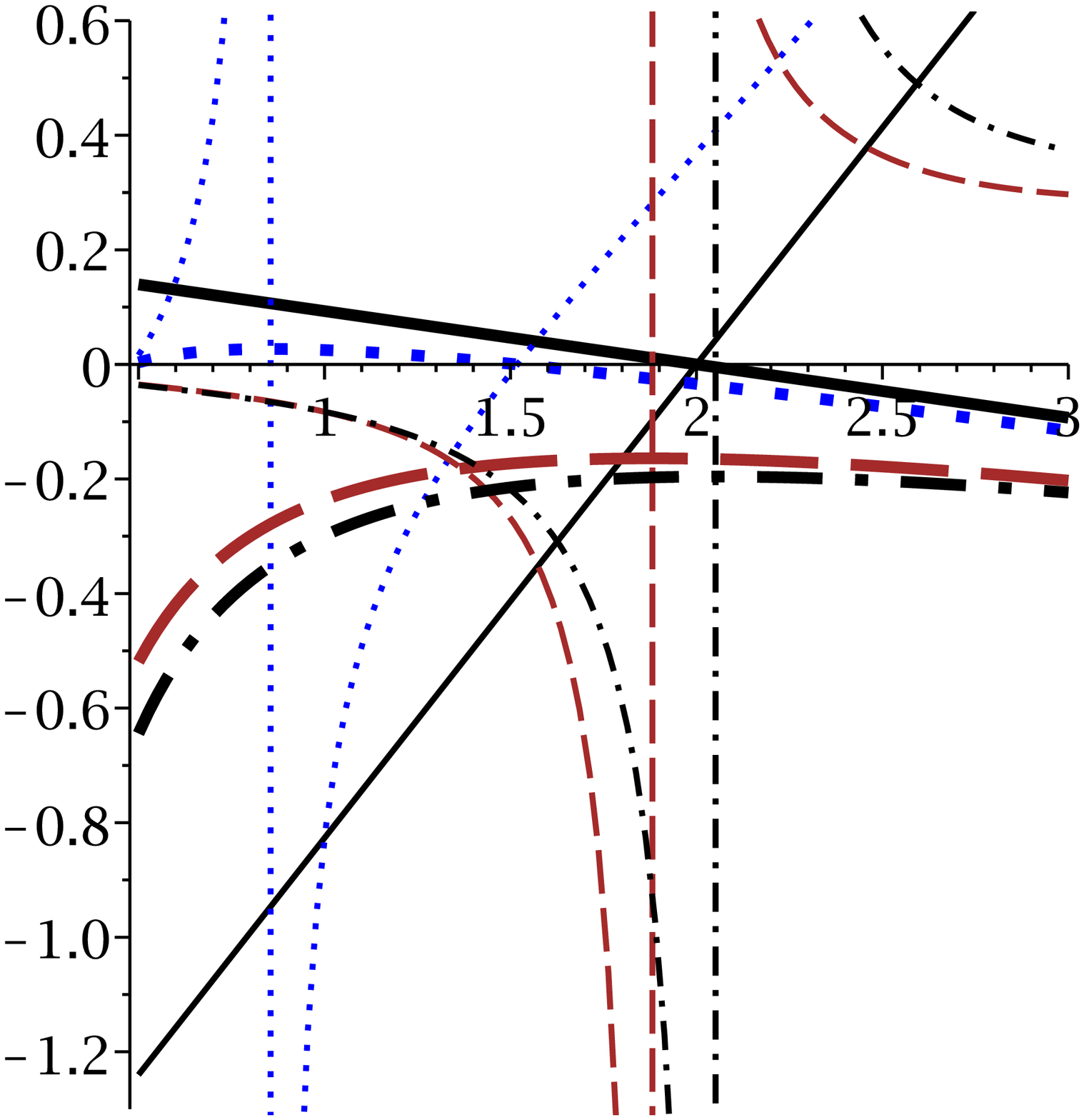}%
\end{array}
$%
\caption{For different scales: $C_{Q}$ and $T$ (bold lines) versus $r_{+}$
for $G(\protect\varepsilon)=l(\protect\varepsilon)=1$, $c(\protect\varepsilon%
)=c_{1}(\protect\varepsilon)=2$, $g(\protect\varepsilon)=1.9$ and $\Lambda(%
\protect\varepsilon)=1$. \newline
Left panel: $q(\protect\varepsilon)=0.5$, $f(\protect\varepsilon)=0.9$, $m(%
\protect\varepsilon)=0$ (continuous line), $m(\protect\varepsilon)=0.92$
(dotted line) and $m(\protect\varepsilon)=1$ (dashed line). \newline
Middle panel: $q(\protect\varepsilon)=0.5$, $m(\protect\varepsilon)=1$, $f(%
\protect\varepsilon)=0.9$ (continuous line), $f(\protect\varepsilon)=1.07$
(dotted line) and $f(\protect\varepsilon)=1.5$ (dashed line). \newline
Right panel: $f(\protect\varepsilon)=0.9$, $m(\protect\varepsilon)=1$, $q(%
\protect\varepsilon)=0$ (continuous line), $q(\protect\varepsilon)=0.5$
(dotted line), $q(\protect\varepsilon)=1.1$ (dashed line) and $q(\protect%
\varepsilon)=1.2$ (dashed-dotted line).}
\label{Fig6}
\end{figure}

\begin{figure}[tbp]
$%
\begin{array}{ccc}
\epsfxsize=6cm \epsffile{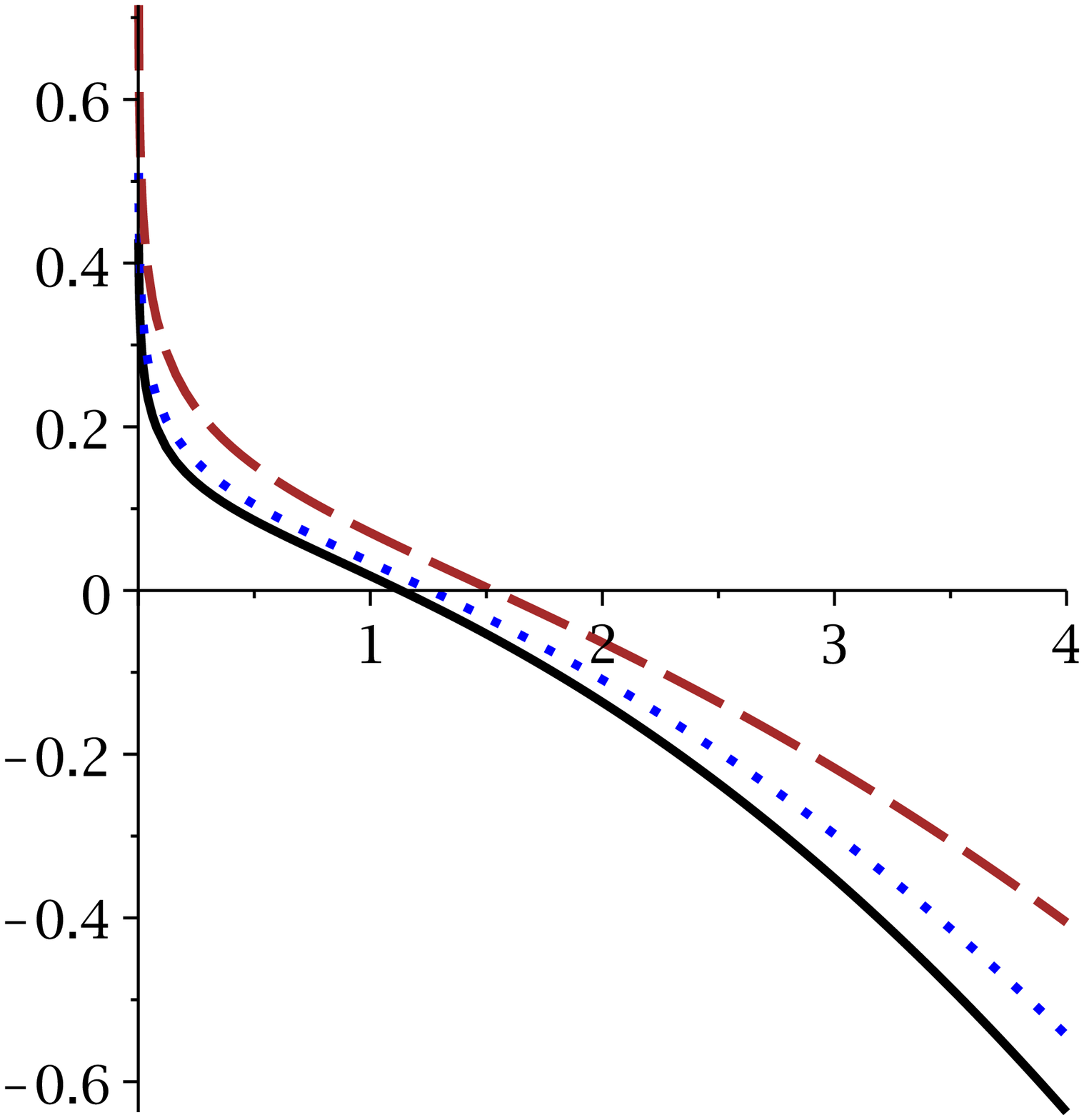} & \epsfxsize=6cm %
\epsffile{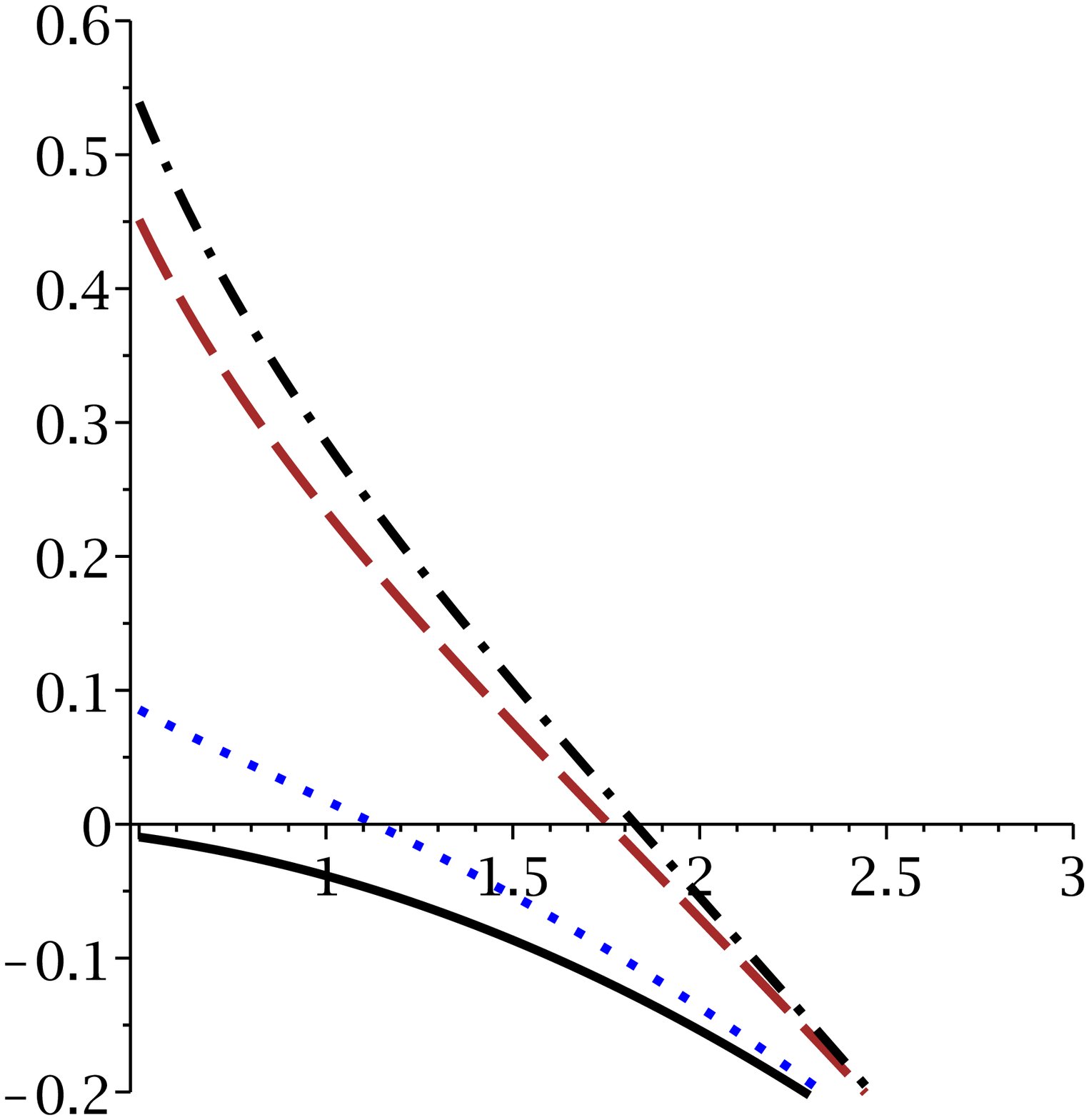} &  \\
\epsfxsize=6cm \epsffile{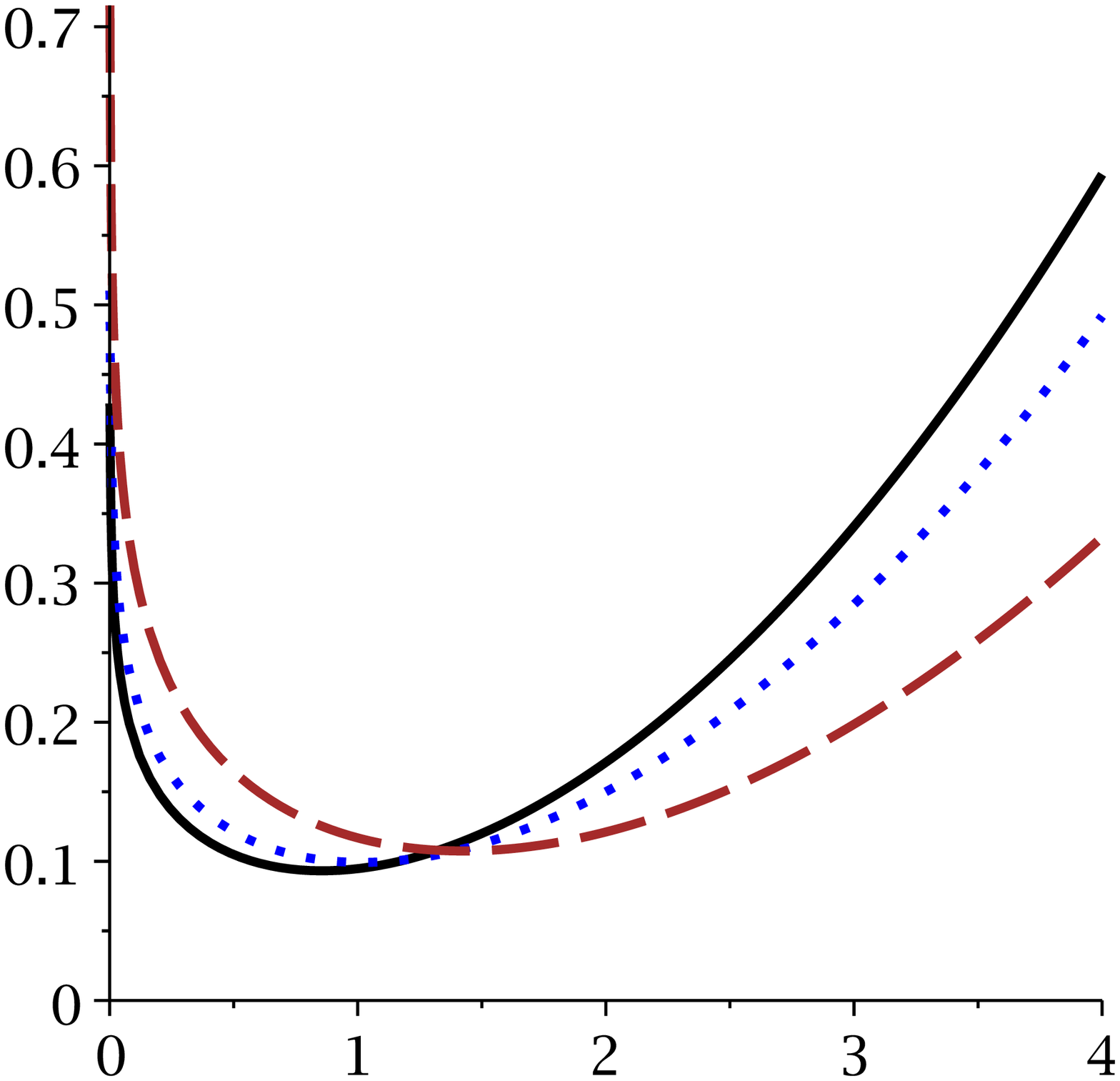} & \epsfxsize=6cm %
\epsffile{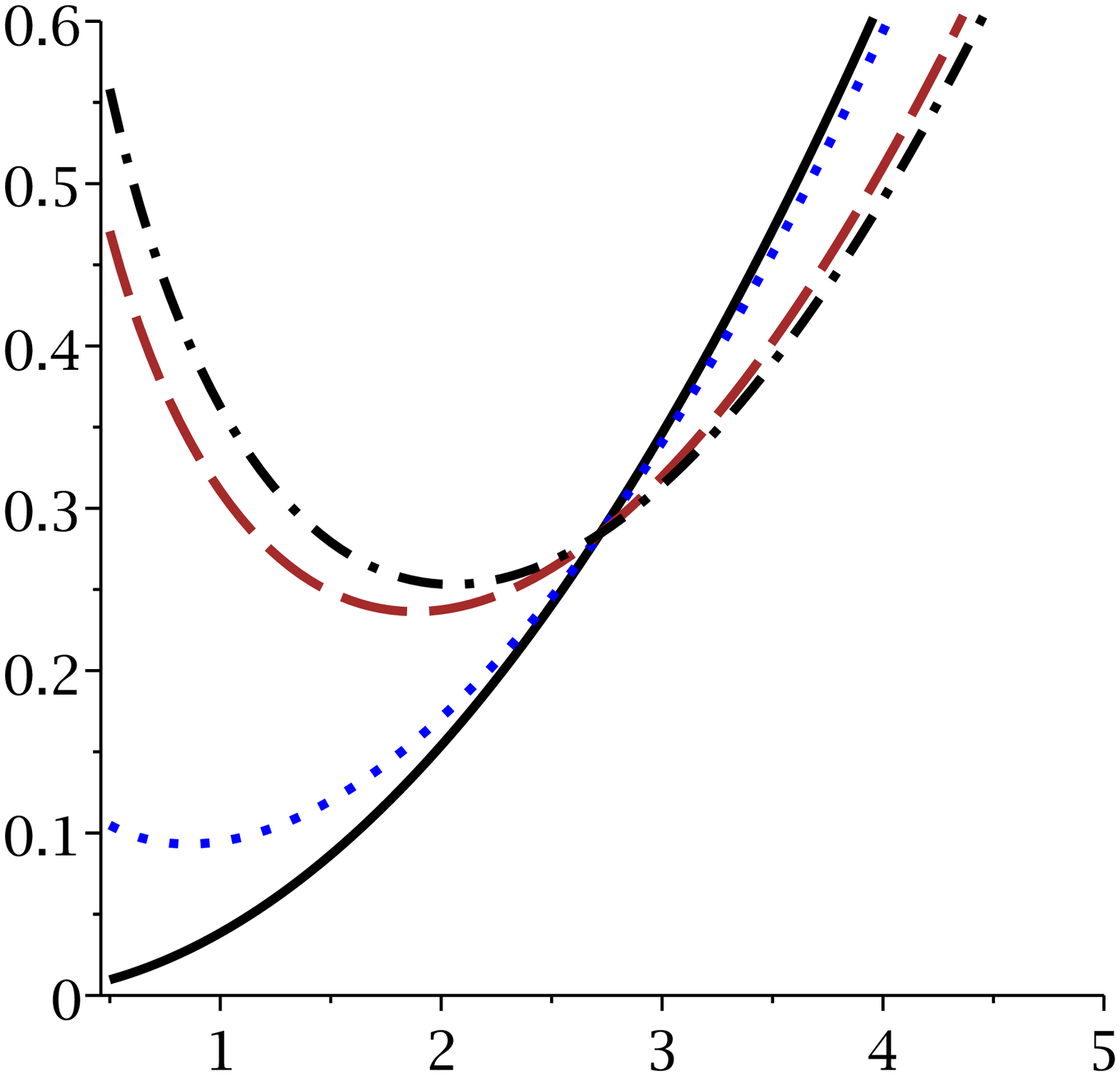} &
\end{array}
$%
\caption{For different scales: $F$ versus $r_{+}$ for $G(\protect\varepsilon%
)=l(\protect\varepsilon)=1$ and $g(\protect\varepsilon)=1.9$. \newline
Left panel: $q(\protect\varepsilon)=0.5$, $f(\protect\varepsilon)=0.9$
(continuous line), $f(\protect\varepsilon)=1.07$ (dotted line) and $f(%
\protect\varepsilon)=1.9$ (dashed line). \newline
Right panel: $f(\protect\varepsilon)=0.9$, $q(\protect\varepsilon)=0$
(continuous line), $q(\protect\varepsilon)=0.5$ (dotted line), $q(\protect%
\varepsilon)=1.1$ (dashed line) and $q(\protect\varepsilon)=1.2$
(dashed-dotted line). \newline
Up panels: $\Lambda(\protect\varepsilon)=-1$; Down panels: $\Lambda(\protect%
\varepsilon)=1$;}
\label{Fig7}
\end{figure}

In order to complete our discussion regarding thermodynamical structure of
these black holes, we have plotted a series of the diagrams for the
mass/internal energy (Figs. \ref{Fig3} and \ref{Fig4}), the temperature
(Figs. \ref{Fig5} and \ref{Fig6}), the heat capacity (Figs. \ref{Fig5} and %
\ref{Fig6}) and the free energy (Fig. \ref{Fig7}) for two cases of dS and
adS spacetime. In Ref. \cite{Mamasani}, it was shown that in order to remove
existence of ensemble dependency, $l(\varepsilon )$ which was inserted for
the sake of dimensionless argument, should be replaced by following relation
\begin{equation}
\Lambda ( \varepsilon) =\pm \frac{1}{l(\varepsilon )},  \label{Lam}
\end{equation}%
in which positive branch is related to dS spacetime while the opposite is
for AdS solutions. Hereafter, we use Eq. (\ref{Lam}) to plot phase diagrams.

Studying mass/internal energy diagrams for adS black holes shows that
depending on the choices of different parameters, this quantity could have a
minimum. This minimum could have a negative mass/internal energy which
indicates that two roots for this quantity exists with region of negative
mass/internal energy. The exception is for the absence and small values of
electric charge. For these cases, there exists a mass for the black holes in
the limit of vanishing horizon radius (see Fig. \ref{Fig3} for more details).

As for dS case, the mass/internal energy is a decreasing function of the
horizon radius with one root. The only exception is for the absence of
electric charge. In this case, a maximum is formed for the mass/internal
energy with one root (see Fig. \ref{Fig4} for more details). For dS black
holes, the region of positivity of the mass/internal energy is located
before root.

The temperature and heat capacity have the same roots. Before the
root, for adS black holes, temperature and heat capacity are
negative and solutions are non-physical. Whereas, after the root,
both temperature and heat capacity are positive valued and
solutions are physical and enjoy thermally stability.
Interestingly, in the absence of electric charge, temperature and
heat capacity are positive and non-zero for vanishing horizon
radius, which confirms our earlier discussion regarding the
existence of remnant for these quantities (see Fig. \ref{Fig5} for
more details).

For dS black holes, the behaviors of temperature and heat capacity are
completely different. Here, the temperature has one maximum. If the maximum
is located at a positive temperature, the heat capacity (and temperature)
will have two roots. Between these roots, only positive temperature hence,
the physical solutions exists. Otherwise, temperature is negative and the
solutions are non-physical. At the maximum of temperature, the heat capacity
acquires divergency which marks a phase transition point. The phase
transition is between large black holes to small ones. This shows that
thermally stable black holes only exist between smaller root and divergence
point. The only exception for the presence of maximum in temperature is the
absence of electric charge. In this case, temperature is a decreasing
function of the horizon radius with one root. In positive region of the
temperature, the heat capacity is negative. Therefore, for this case, the
solutions are thermally unstable (see Fig. \ref{Fig6} for more details).

Finally, for adS case, the free energy is only a decreasing function of the
horizon radius. The only exception for this case is for the absence of
electric charge, where the free energy is negative valued without any root
(see up panels of Fig. \ref{Fig5} for more details). For the dS case, a
minimum is formed for the free energy. This minimum represents the point in
which black holes go under second order phase transition. The only exception
for this case is for the absence of electric charge. In this case, the free
energy is an increasing function of the horizon radius (see down panels of
Fig. \ref{Fig7} for more details).

\section{Geometrical thermodynamics}

In this section, we employ GTs approach to investigate thermodynamical
properties of the black holes by using the so-called HPEM metric. Applying
GTs approach, we can extract some information regarding thermodynamical
behavior of the system by studying the Ricci scalar of constructed phase
space. In this method, the phase transition and bound points should be
represented as divergencies of the Ricci scalar. Recent studies in the
context of the GTs approaches for the black hole thermodynamics have shown
that the Ricci scalars of Weinhold, Ruppeiner and Quevedo metrics may lead
to extra divergencies which are not matched with the bound points and the
phase transitions \cite{HPEMI,HPEMII,HPEMIII,HPEMIV}. In other words, there
were cases of mismatch between divergencies of the Ricci scalar and the
mentioned points (bound and phase transition points), and also existence of
extra divergency unrelated to these points were reported \cite%
{HPEMI,HPEMII,HPEMIII,HPEMIV}. In order to overcome the shortcomings of the
mentioned methods (Weinhold, Ruppeiner and Quevedo metrics), the HPEM method
was introduced and it was shown that the specific structure of this metric
provides satisfactory results regarding GTs of different classes of the
black holes. In addition, this metric contains information which enables one
to determine the type of divergencies to distinguish the divergencies
related to the bound points and those correspond to the phase transition
points.

The HPEM metric of a charged black hole is in the following form
\begin{equation}
ds^{2}=S\frac{M_{S}}{M_{QQ}^{3}}\left( -M_{SS}dS^{2}+M_{QQ}dQ^{2}\right) ,
\end{equation}%
in which $M_{X}=\partial M/\partial X$ and $M_{XX}=\partial ^{2}M/\partial
X^{2}$. The denominator of Ricci scalar of this phase space is \cite{HPEMI}
\begin{eqnarray}
\text{denominator}(\mathcal{R}) &=&2S^{3}M_{SS}^{2}M_{S}^{3},  \nonumber \\
&&  \nonumber \\
&=&\frac{\pi ^{10}G(\varepsilon )^{6}f(\varepsilon )}{128g(\varepsilon )^{7}}%
\left[ \Lambda (\varepsilon )r_{+}^{2}-f(\varepsilon )^{2}g(\varepsilon
)^{2}G(\varepsilon )q(\varepsilon )^{2}\right] ^{2}  \nonumber \\
&&  \nonumber \\
&&\left. \times \left[ 2\Lambda (\varepsilon )r_{+}^{2}-m(\varepsilon
)^{2}c(\varepsilon )c_{1}(\varepsilon )r_{+}+2f(\varepsilon
)^{2}g(\varepsilon )^{2}G(\varepsilon )q(\varepsilon )^{2}\right]
^{3}\right. ,
\end{eqnarray}%
using Eq. (\ref{heat}), we can rewrite the above equation in the following
form
\begin{equation}
\text{denominator}(\mathcal{R})=\frac{\pi ^{7}G(\varepsilon
)^{6}f(\varepsilon )}{2048r_{+}^{3}g(\varepsilon )^{9}}\left[ \text{%
denominator}\left( C_{Q}\right) \right] ^{2}\times \left[ \text{numerator}%
\left( C_{Q}\right) \right] ^{3}.  \label{denomR}
\end{equation}

\begin{figure}[tbp]
$%
\begin{array}{ccc}
\epsfxsize=5cm \epsffile{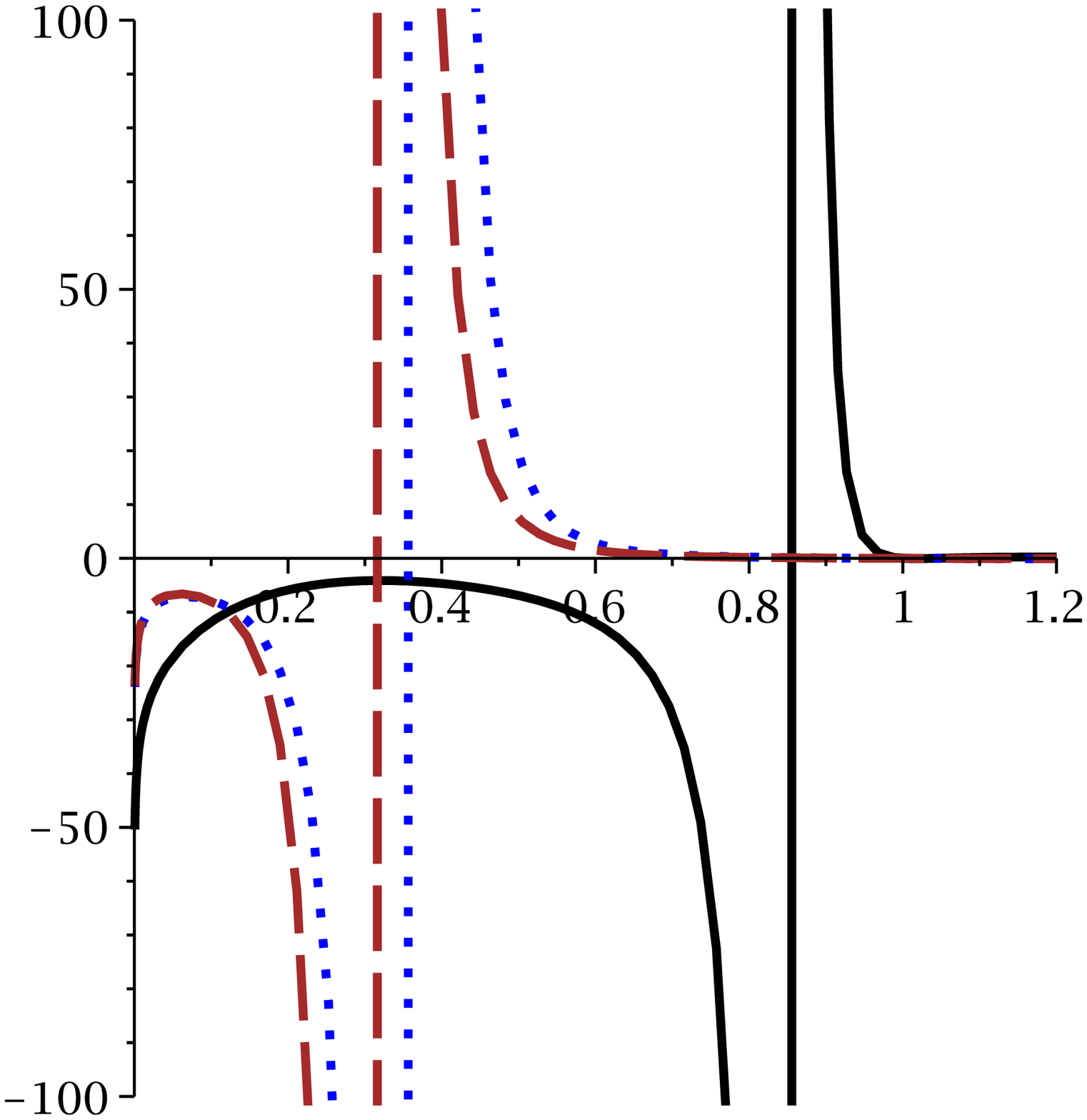} & \epsfxsize=5cm %
\epsffile{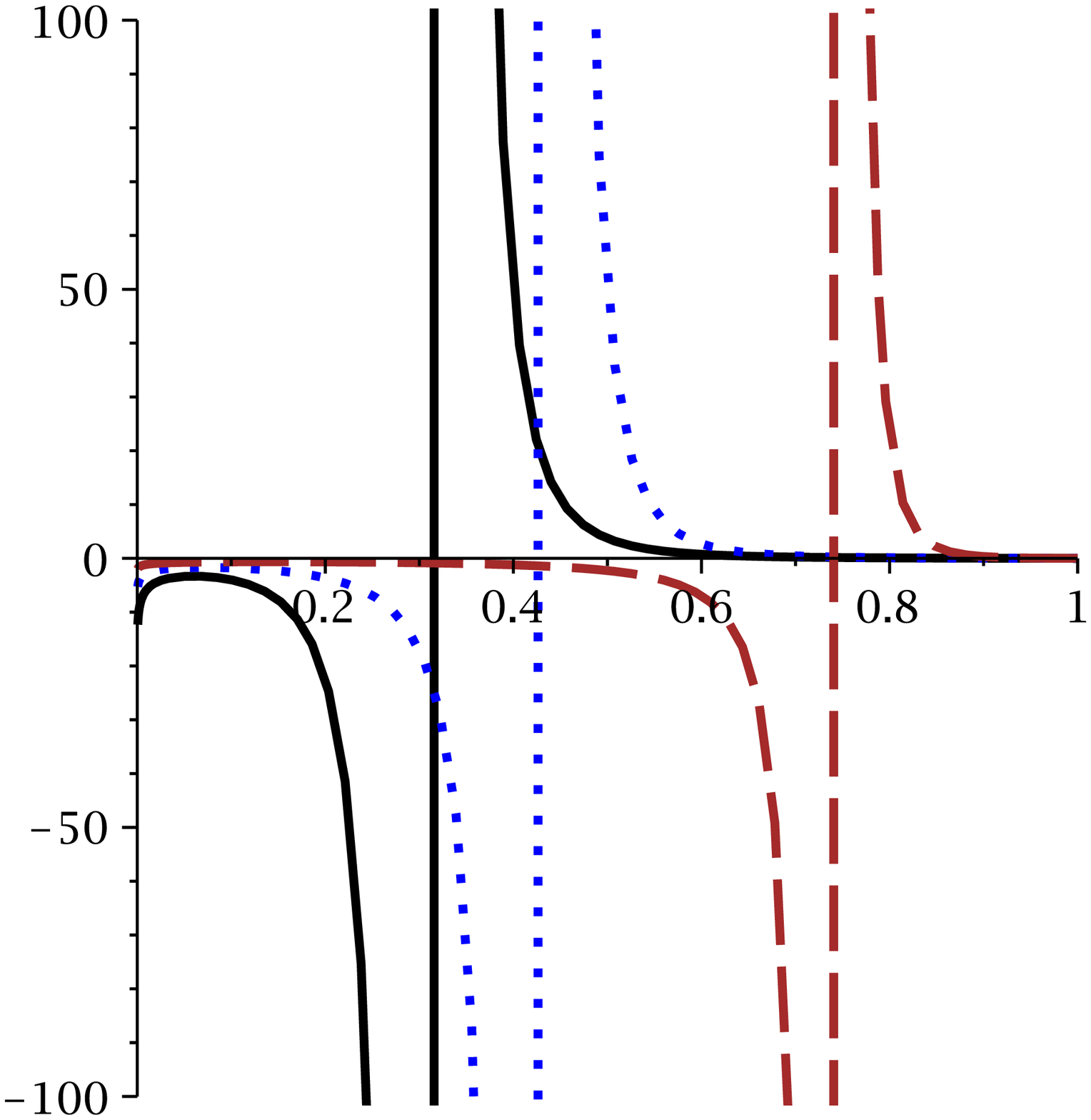} & \epsfxsize=5cm %
\epsffile{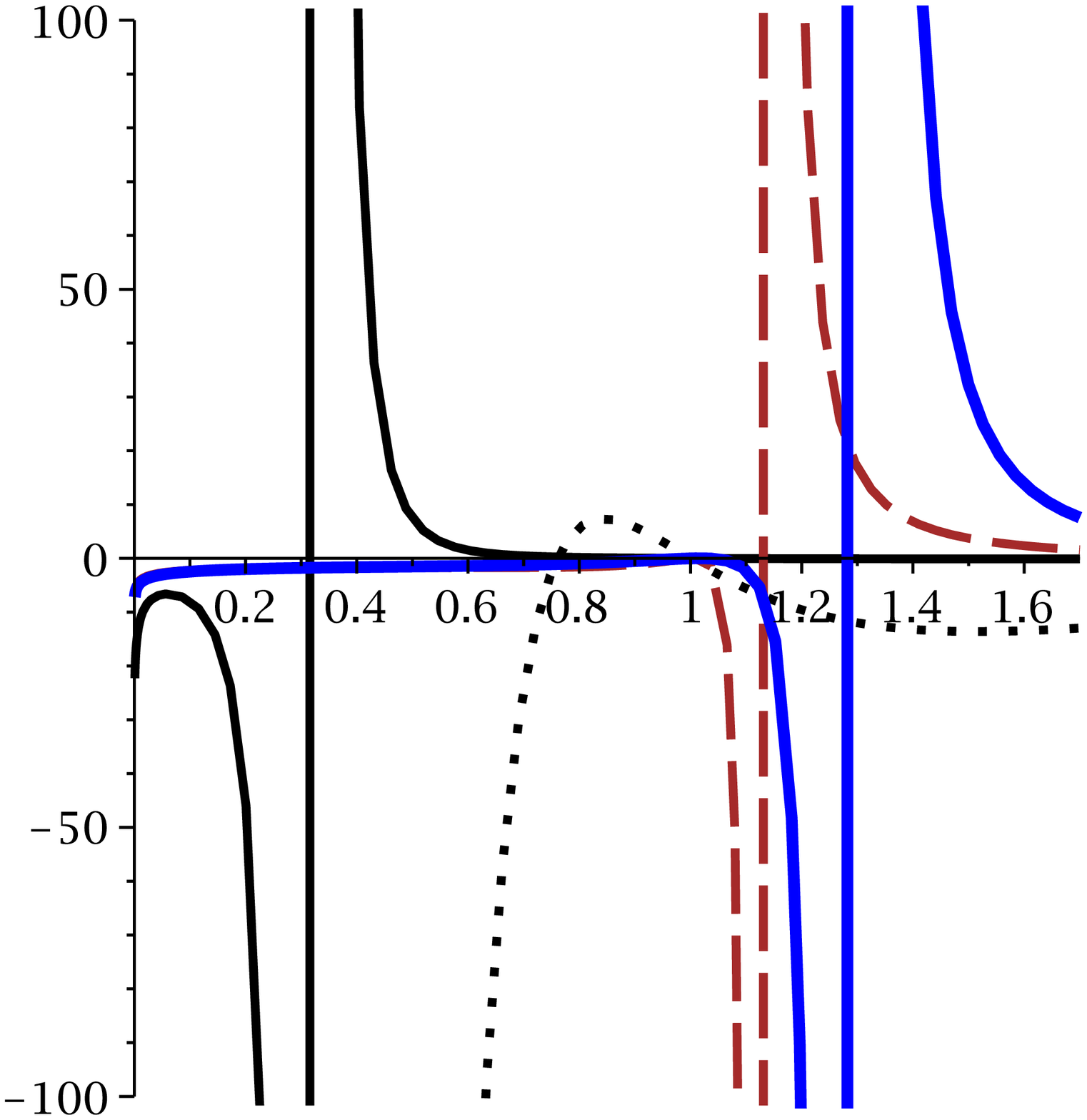}%
\end{array}
$%
\caption{For different scales: $R$ (Ricci scalar) versus $r_{+}$ for $G(%
\protect\varepsilon )=l(\protect\varepsilon )=1$, $c(\protect\varepsilon %
)=c_{1}(\protect\varepsilon )=2$, $g(\protect\varepsilon )=1.9$ and $\Lambda
(\protect\varepsilon )=-1$. \newline
Left panel: $q(\protect\varepsilon )=0.5$, $f(\protect\varepsilon )=0.9$, $m(%
\protect\varepsilon )=0$ (continuous line), $m(\protect\varepsilon )=0.92$
(dotted line) and $m(\protect\varepsilon )=1$ (dashed line). \newline
Middle panel: $q(\protect\varepsilon )=0.5$, $m(\protect\varepsilon )=1$, $f(%
\protect\varepsilon )=0.9$ (continuous line), $f(\protect\varepsilon )=1.07$
(dotted line) and $f(\protect\varepsilon )=1.5$ (dashed line). \newline
Right panel: $f(\protect\varepsilon )=0.9$, $m(\protect\varepsilon )=1$, $q(%
\protect\varepsilon )=0$ (continuous line), $q(\protect\varepsilon )=0.5$
(dotted line), $q(\protect\varepsilon )=1.1$ (dashed line) and $q(\protect%
\varepsilon )=1.2$ (dashed-dotted line).}
\label{Fig8}
\end{figure}

\begin{figure}[tbp]
$%
\begin{array}{ccc}
\epsfxsize=5cm \epsffile{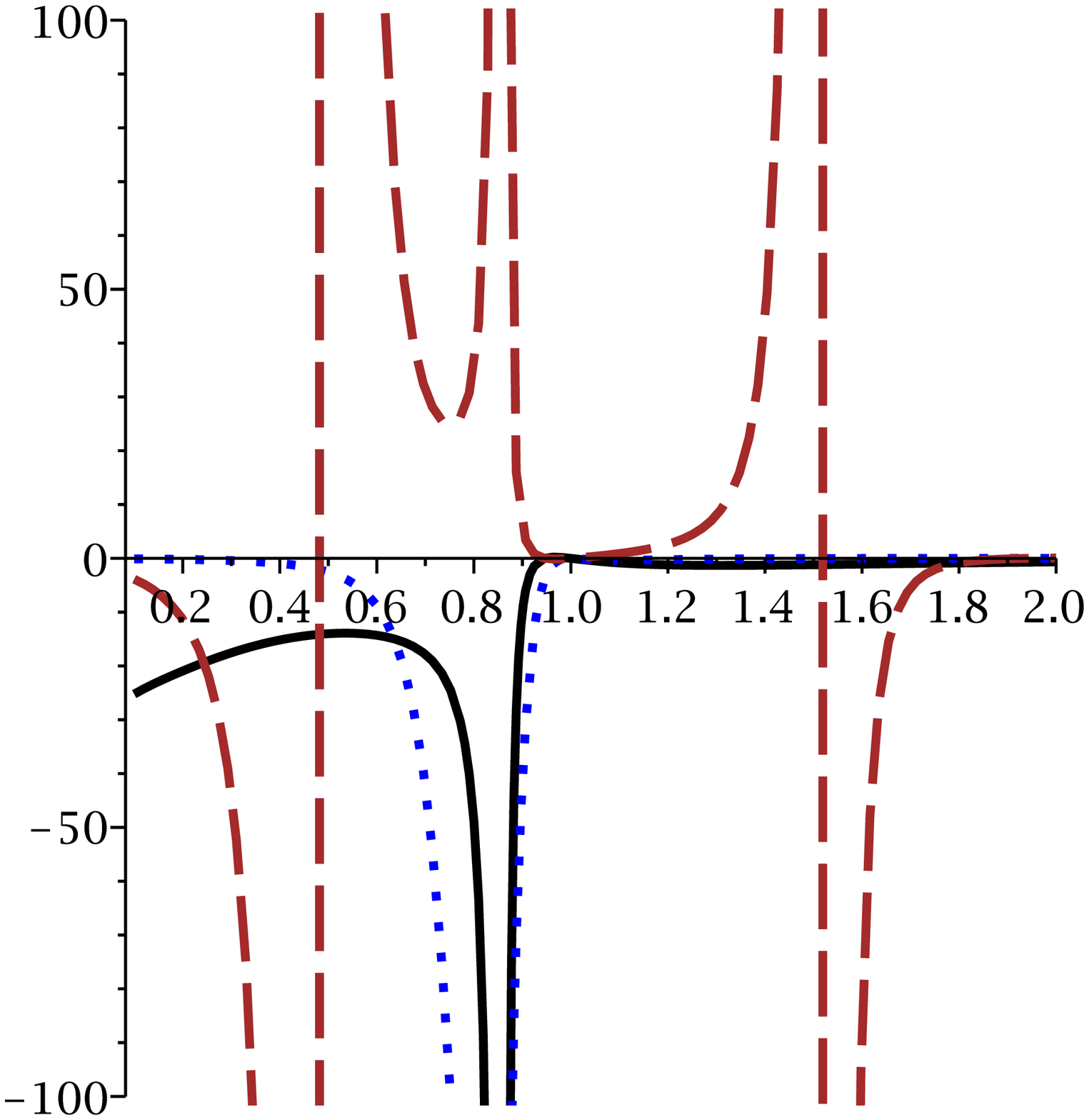} & \epsfxsize=5cm %
\epsffile{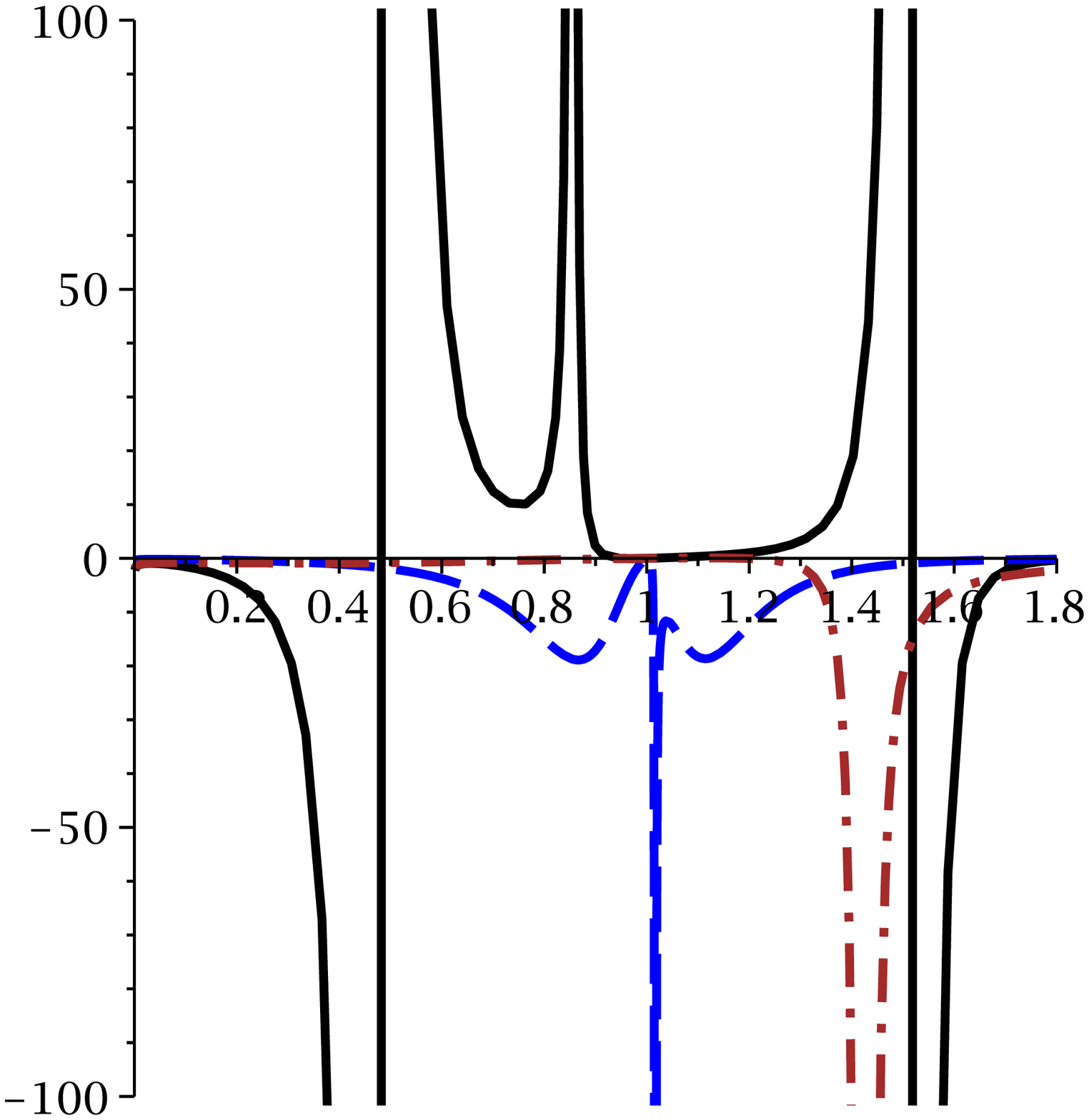} & \epsfxsize=5cm %
\epsffile{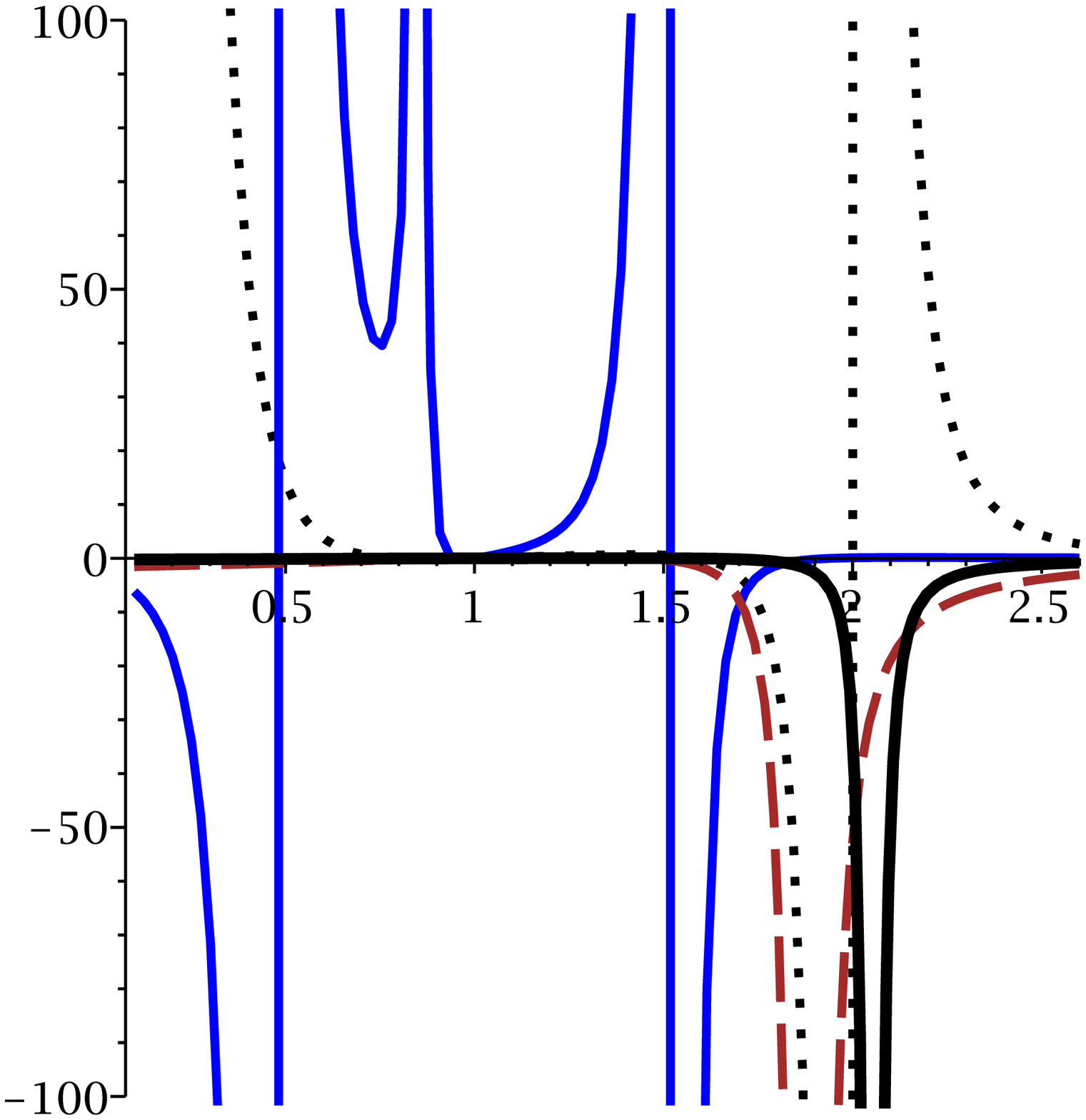}%
\end{array}
$%
\caption{For different scales: $R$ (Ricci scalar) versus $r_{+}$ for $G(%
\protect\varepsilon)=l(\protect\varepsilon)=1$, $c(\protect\varepsilon%
)=c_{1}(\protect\varepsilon)=2$, $g(\protect\varepsilon)=1.9$ and $\Lambda(%
\protect\varepsilon)=-1$. \newline
Left panel: $q(\protect\varepsilon)=0.5$, $f(\protect\varepsilon)=0.9$, $m(%
\protect\varepsilon)=0$ (continuous line), $m(\protect\varepsilon)=0.92$
(dotted line) and $m(\protect\varepsilon)=1$ (dashed line). \newline
Middle panel: $q(\protect\varepsilon)=0.5$, $m(\protect\varepsilon)=1$, $f(%
\protect\varepsilon)=0.9$ (continuous line), $f(\protect\varepsilon)=1.07$
(dotted line) and $f(\protect\varepsilon)=1.5$ (dashed line). \newline
Right panel: $f(\protect\varepsilon)=0.9$, $m(\protect\varepsilon)=1$, $q(%
\protect\varepsilon)=0$ (continuous line), $q(\protect\varepsilon)=0.5$
(dotted line), $q(\protect\varepsilon)=1.1$ (dashed line) and $q(\protect%
\varepsilon)=1.2$ (dashed-dotted line).}
\label{Fig9}
\end{figure}

Comparing Eq. (\ref{denomR}) with the obtained heat capacity
(\ref{heat}), it is evident that bound points and phase transition
points of the heat capacity are matched with divergencies of the
Ricci scalar of the HPEM metric for different parameters. In order
to have better picture, we employ HPEM metric and present table
\ref{tab1} and plot following two diagrams (Figs. \ref{Fig8} and
\ref{Fig9}). In the table \ref{tab1}, $R_{\infty }$ and $C_{\infty
}$ are, respectively, divergencies of the Ricci scalar and heat
capacity. Also, $C_{0}$ and $T_{0}$ are the roots of heat capacity
and temperature, respectively. Comparison between Figs. \ref{Fig8}
and \ref{Fig9}
with plotted diagrams in Figs. \ref{Fig5} and \ref{Fig6} (or see the table %
\ref{tab1}, for more details), shows that all the bound and phase transition
points are matched with divergencies of the Ricci scalar of the HPEM metric
for different parameters. These coincidences between the divergencies of the
Ricci scalar of HPEM metric with the bound and the phase transition points
of heat capacity and the temperature, confirm the validity of results of
this thermodynamical metric. So, one can use this method as an independent
approach regarding studying thermodynamical properties of the black holes.
Another interesting property of HPEM metric is related to the sign of Ricci
scalar of HPEM metric around the bound and the phase transition points which
depends on the type of point. As one can see, the sign of Ricci scalar
around the bound point changes, while for the phase transition point, the
sign of Ricci scalar of HPEM metric does not change (see Figs. \ref{Fig8}
and \ref{Fig9}, for more details). Therefore, by studying the sign of Ricci
scalar of HPEM metric, we can characterize the type of divergencies. On the
other hand, in GTs, the sign of Ricci scalar determines whether system has
attractive (for negative sign) or repulsive (for positive sign) interaction
around the bound and phase transition point. Here, we see that before the
bound point, system has repulsive interaction and by crossing the bound
point, the interaction is changed into attractive (see Figs. \ref{Fig8} and %
\ref{Fig9}, for more details). On the other hand, for the phase transition
point, the sign of Ricci scalar is positive (see Figs. \ref{Fig8} and \ref%
{Fig9}, for more details). It is notable that, we can not extract these
information by using the temperature and the heat capacity of system.
Therefore, we see that employing the HPEM metric provides extra information
regarding the nature of interactions around the bound and the phase
transition points.


\begin{table}[tbp]
\begin{center}
\begin{tabular}{cccccccc}
\hline\hline
$\Lambda(\varepsilon)$ & $f(\varepsilon)$ & $m(\varepsilon)$ & $%
q(\varepsilon)$ & $R_{\infty}$ & $C_{\infty}$ & $C_{0}$ & $T_{0}$ \\
\hline\hline
$-1$ & $0.9$ & $0$ & $0.5$ & $0.85500$ & $-$ & $0.85500$ & $0.85500$ \\
\hline
$-1$ & $0.9$ & $0.92$ & $0.5$ & $0.35668$ & $-$ & $0.35668$ & $0.35668$ \\
\hline
$-1$ & $0.9$ & $1$ & $0.5$ & $0.31568$ & $-$ & $0.31568$ & $0.31568$ \\
\hline\hline
$-1$ & $0.9$ & $1$ & $0.5$ & $0.31568$ & $-$ & $0.31568$ & $0.31568$ \\
\hline
$-1$ & $1.07$ & $1$ & $0.5$ & $0.42592$ & $-$ & $0.42592$ & $0.42592$ \\
\hline
$-1$ & $1.5$ & $1$ & $0.5$ & $0.74086$ & $-$ & $0.74086$ & $0.74086$ \\
\hline\hline
$-1$ & $0.9$ & $1$ & $0$ & $-$ & $-$ & $-$ & $-$ \\ \hline
$-1$ & $0.9$ & $1$ & $0.5$ & $0.31568$ & $-$ & $0.31568$ & $0.31568$ \\
\hline
$-1$ & $0.9$ & $1$ & $1.1$ & $1.13029$ & $-$ & $1.13029$ & $1.13029$ \\
\hline
$-1$ & $0.9$ & $1$ & $1.2$ & $1.28269$ & $-$ & $1.28269$ & $1.28269$ \\
\hline\hline
$1$ & $0.9$ & $0$ & $0.5$ & $0.85500$ & $0.85500$ & $-$ & $-$ \\ \hline
$1$ & $0.9$ & $0.92$ & $0.5$ & $0.85500$ & $0.85500$ & $-$ & $-$ \\ \hline
$1$ & $0.9$ & $1$ & $0.5$ & $(0.48137,0.85500,1.51862)$ & $0.85500$ & $%
(0.48137,1.51862)$ & $(0.48137,1.51862)$ \\ \hline\hline
$1$ & $0.9$ & $1$ & $0.5$ & $(0.48137,0.85500,1.51862)$ & $0.85500$ & $%
(0.48137,1.51862)$ & $(0.48137,1.51862)$ \\ \hline
$1$ & $1.07$ & $1$ & $0.5$ & $1.01650$ & $1.01650$ & $-$ & $-$ \\ \hline
$1$ & $1.5$ & $1$ & $0.5$ & $1.42500$ & $1.42500$ & $-$ & $-$ \\ \hline\hline
$1$ & $0.9$ & $1$ & $0$ & $2.00000$ & $-$ & $2.00000$ & $2.00000$ \\ \hline
$1$ & $0.9$ & $1$ & $0.5$ & $(0.48137,0.85500,1.51862)$ & $0.85500$ & $%
(0.48137,1.51862)$ & $(0.48137,1.51862)$ \\ \hline
$1$ & $0.9$ & $1$ & $1.1$ & $1.88100$ & $1.88100$ & $-$ & $-$ \\ \hline
$1$ & $0.9$ & $1$ & $1.2$ & $2.05200$ & $2.05200$ & $-$ & $-$ \\ \hline\hline
&  &  &  &  &  &  &
\end{tabular}%
\end{center}
\caption{Roots and divergencies of the Ricci scalar, heat capacity and
temperature for $G(\protect\varepsilon)=l(\protect\varepsilon)=1$, $g(%
\protect\varepsilon)=1.9$ and $c(\protect\varepsilon)=c_{1}(\protect%
\varepsilon)=2$.}
\label{tab1}
\end{table}


\section{Conclusions}

In this paper, we have considered the three dimensional black holes in the
presence of massive gravity's rainbow. The solutions were extracted and the
effects of massive gravity and gravity's rainbow on the geometrical
structure of the black holes were studied.

Furthermore, the conserved and thermodynamical properties of these black
holes were extracted in both canonical and grand canonical ensembles. It was
shown that the existence of massive gravity and generalization of the
gravity's rainbow modify thermodynamical quantities of the black holes. The
effects of these generalizations were studied for different thermodynamical
quantities.

It was shown that by suitable choices of different parameters, it is
possible to obtain a minimum for the mass. This minimum could be located at
negative values of the mass/internal energy which leads to the existence of
a region of negativity for the mass/ineternal energy and two roots.
Existence of the negative mass/internal energy is not valid in the classical
thermodynamics of the black holes. Therefore, one can conclude that for this
region of the negative mass/internal energy, black hole solutions do not
exist. This puts a limitation on the values that different parameters can
acquire.

Next, the temperature was taken into account and it was pointed out that one
can acquire a root for this quantity. This root separates the physical
solutions with positive temperature from non-physical ones with negative
temperature. Remarkably, we have observed that the existence of root was
restricted with satisfaction of a condition which was depending on the
values of different parameters. In addition, it was shown that for a
specific limit, there may exist a remnant of temperature for these black
holes. In other words, after the evaporation of black holes, there will be
traces of existence of these black holes in form of fluctuation in
temperature of the spacetime.

Subsequently, the heat capacity was investigated and the possibility of
divergence point was pointed out. This divergency marks the phase transition
point for these black holes. The existence of real valued positive
divergence point depends on the spacetime being dS or adS. The asymptotic
and high energy limits of the heat capacity were studied as well and it was
shown that these limits were governed by the gravity's rainbow and massive
gravity generalizations.

After that, the free energy was studied and its root and divergence point
were extracted. It was shown that the extremum of free energy and divergence
point of the heat capacity are matched. The resulting critical horizon
radius was used to obtain the critical temperature, the mass/internal energy
and the free energy.

At last, we have used HPEM metric in the context GTs in order to study
thermodynamical structure of these black holes. It was shown that this
metric can describe the (non)physical and phase transition points of these
black holes. Besides, by studying the sign of HPEM metric around the bound
and phase transition points, it was possible to distinguish repulsive and
attractive interactions of thermodynamical system.

The rainbow functions of the metric are originated form quantum
corrections. Study conducted in this paper showed significant
modifications in thermodynamics of the black holes with
consideration of the such corrections. In other words, it was
shown that in semi-classical/quantum regime, thermodynamics of the
black holes would be modified into a level which differers from
classical case. We observed that different orders of the rainbow
functions affect the high energy and asymptotical behaviors of the
solutions and their leading terms. On the other hand, we found
that the highest contribution of massive gravitons could be
observed in medium black holes. The only case, in which the
effects of massive gravity could be observed for small (large)
black holes was in the case of vanishing electric charge
(cosmological constant). Obtained results here could be employed
to study lattice like behavior in context of AdS/QCD
correspondence. In addition, it is possible to employ the results
of this paper to investigate the entropy spectrum and quasinormal
modes. In fact, it would be interesting to investigate the effects
of energy functions on these quantities. The results of this paper
could also be employed in the context of AdS/CFT correspondence
and central charges, specially considering the energy dependency
of the constants and their effects on calculation of the central
charges.

\begin{acknowledgements}
We would like to thank the referee for his/her insightful
comments. We also thank Shiraz University Research Council. This
work has been supported financially by the Research Institute for
Astronomy and Astrophysics of Maragha, Iran.
\end{acknowledgements}

\end{document}